\documentclass{emulateapj}

\begin{document}

\shortauthors{Luhman}
\shorttitle{Census of Taurus and Neighboring Associations}

\title{A Census of the Taurus Star-forming Region and Neighboring Associations with Gaia\altaffilmark{1}}

\author{K. L. Luhman\altaffilmark{2,3}}

\altaffiltext{1}
{Based on observations made with the Gaia mission, the Two Micron All
Sky Survey, the Wide-field Infrared Survey Explorer, the LAMOST survey,
the Sloan Digital Sky Survey IV, the NASA Infrared Telescope Facility,
and Gemini Observatory.}

\altaffiltext{2}{Department of Astronomy and Astrophysics,
The Pennsylvania State University, University Park, PA 16802, USA;
kll207@psu.edu}

\altaffiltext{3}{Center for Exoplanets and Habitable Worlds, The
Pennsylvania State University, University Park, PA 16802, USA}

\begin{abstract}

I have used high-precision photometry and astrometry from the third data
release of Gaia (DR3) to perform a survey for members of the Taurus
star-forming region and young associations in its vicinity. 
This work has produced a new catalog of 532 adopted members of Taurus, which 
has only minor changes relative to the previous catalog from \citet{esp19}.
I have used the Gaia astrometry to divide the Taurus members into
13 groups that have distinct kinematics. 
Meanwhile, I have identified 1378 candidate members of seven associations
near Taurus. All of these associations have histograms of spectral types
that peak near M5 ($\sim0.15$~$M_\odot$), resembling other young populations
in the solar neighborhood.
For the Taurus groups and neighboring associations, I have estimated
ages from their sequences of low-mass stars in Hertzsprung-Russell diagrams.
Most of the Taurus groups have median ages of $\sim$1--3~Myr while
the associations have ages ranging from 13 to 56~Myr.
I have used mid-infrared photometry from the Wide-field Infrared Survey 
Explorer to search for excess emission from circumstellar disks among the 
candidate members of the associations. Disks are detected for 51 stars, 
20 of which are reported for the first time in this work. 
Some recent studies have proposed that samples of older stars ($\gtrsim$10~Myr)
found in the vicinity of Taurus represent a distributed population that
is associated with the Taurus cloud complex. However, I find that most
of those stars have kinematics that are inconsistent with any relationship
with Taurus.

\end{abstract}

\section{Introduction}

The solar neighborhood contains numerous groups, associations, and clusters 
that span a wide range of ages. The youngest populations 
($\lesssim5$~Myr) are associated with their natal molecular clouds and are
located at distances of $>$100~pc.
The Taurus complex of clouds is one of the nearest sites of recent
star formation \citep[$d\sim140$~pc,][references therein]{gal18}.
Roughly a dozen small groups of stars are associated with the clouds in Taurus,
which together have a total of $\sim500$ proposed members \citep{esp19}.
The Taurus groups appear to have roughly similar ages (few Myr) and yet
have a long crossing time (10--20~Myr) because of their large physical extent,
which has been cited as evidence that the formation of molecular clouds
is rapid \citep{bal99,har01b}.

Since the Taurus groups have low stellar density and cover a large area of sky, 
membership surveys based on signatures of youth have been potentially 
subject to contamination from young stars in the field and in
unrelated associations \citep{har91,bri97,dez99}.
It should be possible to greatly reduce such contamination using data
from the Gaia mission \citep{per01,deb12,gaia16b},
which is performing an all-sky survey to measure high-precision photometry,
proper motions, and parallaxes for more than a billion stars as faint as
$G\sim20$ ($\sim0.05$~$M_\odot$ in Taurus).
In addition, kinematic data from Gaia should enable discrimination among
the Taurus groups.
However, recent studies of Taurus with Gaia have produced substantially
different results on both fronts \citep{luh18,gal19,roc20,liu21,ker21,kro21}.

In this paper, I have used the third data release of Gaia 
\citep[DR3,][]{bro21,val22} to perform a census of the Taurus groups and
associations in their vicinity.  I describe the construction of my catalog of 
adopted members of Taurus (Section~\ref{sec:cat}), assign those members to 
individual Taurus groups based on their kinematics, and search for new members 
of the groups that have similar kinematics (Section~\ref{sec:tausearch}).
I then identify candidate members of young associations near Taurus
(Section~\ref{sec:groupsearch}), characterize the properties
of the stellar populations in the Taurus groups and the neighboring
associations (Section~\ref{sec:pops}), and compare the results
of my work to those of recent studies (Section~\ref{sec:compare}).

\section{Catalog of Taurus Members}
\label{sec:cat}

\subsection{Adopted Members}
\label{sec:adopt}

My previous catalog of adopted members of Taurus contained 519 objects
\citep{esp19}.
In that catalog, the components of multiple systems were given separate
entries if they were resolved by Gaia or other wide-field imaging surveys. 
Otherwise, the components appeared together in a single entry
if they were resolved only in a multiplicity survey \citep[e.g.,][]{kon07}.
Among the latter, the following eight companions are resolved by Gaia DR3
and have separate entries in my new catalog:
LkCa~3B, LkCa~7B, XZ~Tau~B, HBC~412~B, IW~Tau~B, JH112~Ab, J2-2041~B, and
a companion to 2MASS J04400174+2556292 (Gaia DR3 148386898125118464).
Gaia DR3 149370651794032640 and 149370651795329536 
have a separation of $0\farcs4$ and are located near the infrared (IR) position
of IRAS~04264+2433. In optical images from the Hubble Space Telescope
(program 9103, K. Stapelfeldt), the first Gaia source is brighter and more 
point-like, so it is adopted as the counterpart to IRAS~04264+2433. The second 
Gaia source may be nebulosity rather than a star, so it is not included in my 
catalog. I have rejected eight stars from \citet{esp19} based on the analysis 
of Gaia DR3 astrometry in Section~\ref{sec:tausearch}, consisting of BS~Tau~A/B,
HD~283782, Gaia DR3 151851867285622528, 151870352825256576, 
151639902060635520, 157644373715415424, and 155964354307956480.
Meanwhile, I have adopted the following 13 objects as new members based 
on their Gaia kinematics and spectroscopic confirmation of youth 
(Section~\ref{sec:spec}): PW Aur, Gaia DR3 145200960104259200,
3420750548559422592, 3414676232147787136, 3418846267435680512,
3446890376655192832, 3446890411014932224, 180149418232233472,
156207518176564864, 157247965413620224, 156162674425653248, 3420824426291884672,
and 3421359544857244160. My new catalog of Taurus members contains 532 objects.

\subsection{New Spectral Classifications}
\label{sec:spec}

Gaia DR3 3419115132386033280 is a $0\farcs95$ candidate companion 
that has been adopted as a Taurus member in my recent studies but has 
lacked a spectral classification. I obtained a near-IR spectrum of it
using SpeX \citep{ray03} at the NASA Infrared Telescope Facility (IRTF)
on the night of 2020 January 2. The spectrum was reduced with
the Spextool package \citep{cus04}, which included correction of 
telluric absorption \citep{vac03}. A spectral type was measured through
comparison to standard spectra derived from optically-classified young
stars \citep{luh17}.

Thirteen of the candidate members of Taurus identified with Gaia DR3 in
Section~\ref{sec:tausearch} have spectroscopy available for verifying their
youth. Two of those objects, Gaia DR3 3414676232147787136
and PSO J079.3986+26.2455, have been spectroscopically confirmed as young
in previous work \citep{sle06,zha18}. I have measured a new spectral type
from the previous data for the latter using the standards from \citet{luh17}.
The remaining 11 stars have been observed through the
Large Sky Area Multi-Object Fiber Spectroscopic Telescope survey
\citep[LAMOST;][]{cui12,zha12} or with the GoldCam spectrograph at the
Kitt Peak National Observatory (KPNO) 2.1~m telescope 
(program 2010B-0530, S. Takita).
The LAMOST data were taken from the seventh data release (DR7) of the survey.
All of these stars exhibit evidence of youth in their spectra.
I measured their spectral types through comparison
to field dwarf standards for $<$M5 \citep{hen94,kir91,kir97} and
averages of dwarf and giant standards for $\geq$M5 \citep{luh97,luh99}.

The new spectral classifications are provided in Table~\ref{tab:spec}.

\subsection{Compilation of Data}
\label{sec:compilation}

My catalog of adopted Taurus members is presented in Table~\ref{tab:mem},
which includes source names from Gaia DR3, the Point Source Catalog of the 
Two Micron All Sky Survey \citep[2MASS,][]{skr03,skr06},
the United Kingdom Infrared Telescope (UKIRT) Infrared Deep
Sky Survey \citep[UKIDSS,][]{law07}, and previous studies; equatorial
coordinates from Gaia DR3, 2MASS, UKIDSS, or the Spitzer Space Telescope; 
measurements of spectral types and the type adopted in this work; 
data from Gaia DR3 that consist of proper motion, parallax, renormalized unit
weight error \citep[RUWE,][]{lin18}\footnote{RUWE provides a measure of 
the goodness of fit for the astrometry.}, and photometry 
in bands at 3300--10500~\AA\ ($G$), 3300--6800~\AA\ ($G_{\rm BP}$),
and 6300-10500~\AA\ ($G_{\rm RP}$);
distance estimate based on the Gaia DR3 parallax \citep{bai21}; 
the most accurate available radial velocity measurement that has an error 
less than 4~km~s$^{-1}$; $UVW$ velocities calculated from the proper motion, 
parallactic distance, and radial velocity \citep{joh87,luh20u}; a flag
indicating whether the object is an outlier in terms of its Gaia astrometry;
and the Taurus group to which the object has been assigned based on
its Gaia astrometry (Section~\ref{sec:assign}).

One of the sources of radial velocities for Taurus members
is the APOGEE-2 program within the Sloan Digital Sky Survey IV 
\citep[SDSS-IV,][]{bla17,maj17,apo17}. The errors in those velocities
are likely underestimated \citep{cot14,tsa22}. When multiple measurements 
are available from APOGEE-2, the average value is adopted in this work.
For those stars, I have included in Table~\ref{tab:mem} the {\tt VSCATTER}
parameter from that survey, which measures the scatter among the multiple
measurements. Large values of {\tt VSCATTER} may indicate the presence of a 
binary, in which case the average velocity may not represent the velocity of
the system. Thus, the radial velocities and corresponding
$UVW$ velocities of such stars should be treated with caution.
The radial velocities adopted from the LAMOST Medium-resolution Survey
were derived from data at multiple epochs in the manner described by
\citet{luh22o}.

Proper motions and parallaxes are available from radio interferometry
\citep{gal18} for four Taurus members that lack such measurements
from Gaia DR3, consisting of LkCa~3~A, V410~Anon~25, XZ~Tau~A, and LkHa332/G1.
An additional member with radio astrometry, the binary system V807~Tau, has
a large parallax error and a poor astrometric fit in Gaia DR3.
The proper motions, parallaxes, and implied distances for these five stars
in Table~\ref{tab:mem} are from \citet{gal18}.

Among the 532 adopted members, 467 have entries in Gaia DR3 and 412 have 
proper motions and parallaxes from Gaia DR3. Those parameters are available 
for 416 members after including the radio measurements from \citet{gal18}.
The number of members with $\sigma_{\pi}<1$~mas is 412, 291 of which
also have radial velocity measurements.  Radial velocities are available
for 330 members, 37 of which lack parallax data.
Most of the sources that lack parallaxes and radial velocities are
close companions, brown dwarfs, or protostars.

The catalog of Taurus members from \citet{esp19} included a compilation of
IR photometry and disk classifications. I have checked the 13 additional stars
adopted as members in this work for evidence of disks using mid-IR photometry
from the AllWISE Source Catalog \citep{cut13a,wri13} produced by 
the Wide-field Infrared Survey Explorer \citep[WISE,][]{wri10}.
Excess emission from disks was identified and classified in the manner
described in previous studies \citep{esp14,esp18}.
The resulting classifications are included in Table~\ref{tab:spec}.
Five stars lack disk emission and are designated as
class~III \citep{lw84,lad87}. The remaining eight stars have disks
that are classified as full, all of which have been detected in previous
studies \citep{reb11,zha18,liu21}.
The disk classification for Gaia DR3 156162674425653248 is tentative because 
it is only slightly resolved from a $5\farcs5$ candidate companion
in the WISE data.

\section{A Search for New Members of Taurus}
\label{sec:tausearch}

\subsection{Survey Strategy}

In \citet{luh18}, I used data from the second data release of Gaia
\citep[DR2,][]{bro18} to search for new members of Taurus based on
positions in color-magnitude diagrams (CMDs) that were indicative of young
stars and based on kinematics that were similar to those of the groups of
known young stars associated with the Taurus clouds. The resulting survey had 
a high level of completeness for spectral types of $\lesssim$M6--M7 at 
low-to-moderate extinctions ($A_J<1$). I have followed the same approach with 
my new survey that uses data from Gaia DR3.
As done in \citet{luh18} and in my studies of other young associations
\citep[e.g.,][]{esp17,luh20u,luh22sc}, I have analyzed
the Gaia astrometry in terms of a ``proper motion offset"
($\Delta\mu_{\alpha,\delta}$), which is defined as the difference between the
observed proper motion of a star and the motion expected at the celestial
coordinates and parallactic distance of the star for a specified
space velocity. By using the proper motion offset, projection effects
are minimized, which is particularly important for regions like Taurus that
cover a large area of sky.
For my survey, the proper motion offsets are calculated relative to the
motions expected for a velocity of $U, V, W = -16, -12, -9$~km~s$^{-1}$,
which approximates the median velocity of Taurus members \citep{luh18}.
For parallactic distances, I adopt the geometric values estimated by
\citet{bai21} from DR3 parallaxes.

\subsection{Assigning Adopted Members to Groups}
\label{sec:assign}

Before searching for new members of Taurus, I need to characterize
the kinematics of the Taurus groups, which requires defining the groups and
the adopted Taurus members that belong to them.
I have divided the adopted members that have measurements of proper motions 
and parallaxes into groups based on their clusters in proper motion offsets,
which I refer to as ``kinematic clusters".
During that process, I have considered the spatial positions of stars in the
following manner. If a given kinematic cluster contains stars that reside
in two spatial clusters that are widely separated (i.e., two groups on
opposite sides of the Taurus complex), they are divided into two groups.
Otherwise, if all of the stars in a kinematic cluster have a continuous spatial
distribution or consist of two adjacent spatial clusters, they are treated as
a single group. An example of the latter is the pair of neighboring clumps of
stars toward the L1495 and B209 clouds, which have similar kinematics.
This analysis has produced 13 groups, which are named after the dark clouds
that they surround or their brightest members (T~Tau, HD~28354).
Among the 416 adopted members with parallax measurements, 412 have been
assigned to the groups, as indicated in Table~\ref{tab:mem}.
The remaining four stars are located in areas where groups overlap spatially
and have sufficiently large astrometric uncertainties that it is unclear to 
which group they belong.
The Taurus members that lack parallaxes have not been assigned to groups,
but for many of them, the identities of their groups are strongly implied
by their locations on the sky (i.e., they are near only a single group).

In Figure~\ref{fig:map1}, I have plotted all adopted members of Taurus
on a map of equatorial coordinates with symbols that indicate their
assigned groups (or the absence of one). An extinction map is included to
show the dark clouds. For each group, the members with parallax measurements
are plotted in diagrams of $G$ versus distance and $\Delta\mu_{\delta}$ versus
$\Delta\mu_{\alpha}$ in Figures~\ref{fig:pp1}--\ref{fig:pp4},
which illustrate the spatial and kinematic clustering of the group members.
Some of the stars in those diagrams have distances or proper motion offsets
that are discrepant relative to their assigned groups, which are flagged
in Table~\ref{tab:mem}. Those stars are retained as members of groups because 
(1) their discrepant astrometry may be explained by poor astrometric fits 
based on their large values of RUWE, the presence of close companions,
or large changes in parallax between DR2 and DR3 or
(2) they appear to have experienced dynamical ejections, which applies to
KPNO~15 and 2MASS J04355209+225503 \citep{luh18}. 
Examples of members in the first category of outliers include
MHO~3 in L1495/B209 (Figure~\ref{fig:pp1}) and GG~Tau~A, HD~28867~A,
HD~30171, and V827~Tau in L1551 (Figure~\ref{fig:pp4}).
All of these stars have large enough values of RUWE \citep[$>$1.6,][]{luh20u}
to indicate that they may have poor astrometric fits. The parallaxes from DR2 
and DR3 for MHO~3 differ significantly, further indicating that its DR3
astrometry may not be reliable. It is not surprising that GG~Tau~A would have a 
poor fit given that it is binary system with a separation just below the
resolution of Gaia \citep[$0\farcs25$,][]{lei91,whi99}.
HD~28867~A is also suspected to have a close companion that is unresolved
in available imaging \citep[$\lesssim0\farcs11$,][]{wal03}.
The astrometry of HD~28867~B, which has a separation of $3\farcs1$
from the primary, agrees well with that of L1551.

Eight of the stars adopted as Taurus members by \citet{esp19} 
have Gaia astrometry that is inconsistent with membership in the Taurus
groups and seem to have reliable astrometry, so they have been rejected and
are absent from Figures~\ref{fig:pp1}--\ref{fig:pp4}. The names of those
stars were provided in Section~\ref{sec:adopt}.

In Section~\ref{sec:compare}, I compare the Taurus groups defined in this
work and to those from previous studies.

\subsection{New Candidate Members of Taurus Groups}

\begin{figure}
\epsscale{1.1}
\plotone{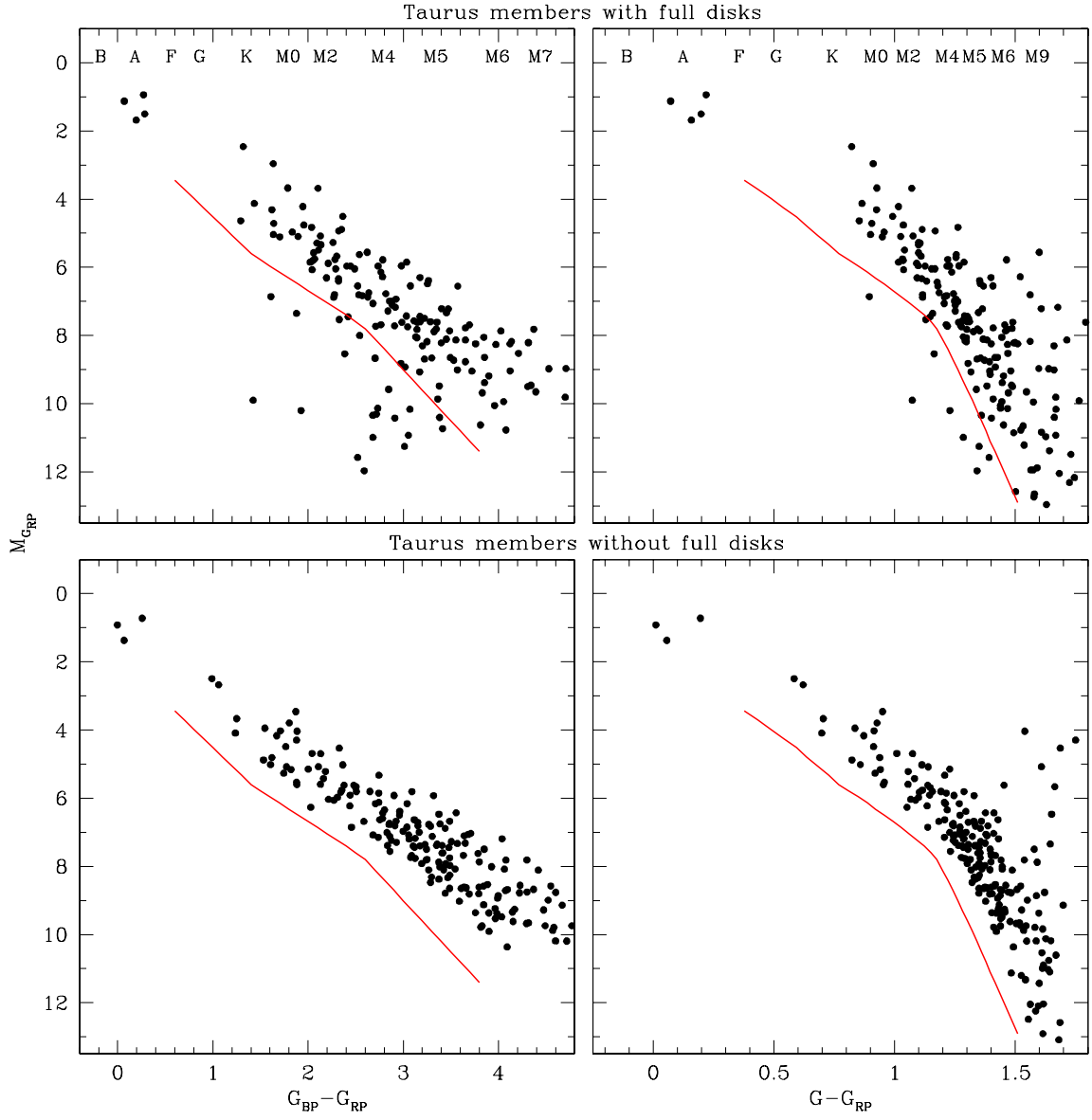}
\caption{
$M_{G_{\rm RP}}$ versus $G_{\rm BP}-G_{\rm RP}$ and $G-G_{\rm RP}$
for members of Taurus that have parallax measurements.
Members that have full disks are shown in the top row and
the remaining stars are plotted in the bottom row.
The boundaries used for selecting candidate members of Sco-Cen by
\citet{luh22sc} are marked (red lines).
For reference, the spectral types that correspond to the colors of young
stars are indicated \citep{luh22sc}.
}
\label{fig:cmd1}
\end{figure}

To illustrate the photometric selection criteria for my search for new
members of Taurus, I have plotted diagrams of $M_{G_{\rm RP}}$ versus
$G_{\rm BP}-G_{\rm RP}$ and $G-G_{\rm RP}$ in Figure~\ref{fig:cmd1}
for the adopted members of Taurus that have parallax measurements.
Separate diagrams are shown for stars with and without full disks.
In each color, the latter form a well-defined sequence except for a few
stars that are very red in $G-G_{\rm RP}$ for their magnitudes, which is
likely a reflection of erroneous photometry in $G$ that is caused by
contamination from close companions \citep{eva18}.
Meanwhile, the sequence of stars with full disks includes an extension to
relatively faint magnitudes at a given color, which can be explained by
short-wavelength emission associated with accretion or scattered light from
an edge-on disk. In Figure~\ref{fig:cmd1}, I have marked the boundaries that
I have previously used for selecting candidate members of populations
in the Scorpius-Centaurus (Sco-Cen) OB association \citep{luh22sc}.
I have applied those boundaries to my survey of Taurus as well.
Thus, my survey is sensitive to stars in the same age range
as found in Sco-Cen ($\lesssim$20~Myr).

I have searched Gaia DR3 for stars that appear above one of the CMD boundaries
in Figure~\ref{fig:cmd1}, do not appear below either CMD boundary,
overlap with the proper motion offsets of a Taurus group, are located
within $\sim5$~pc of a member of that group, and have $\sigma_{\pi}<1$~mas.
Only photometric data with errors less than 0.1~mag were utilized.
Given that disk-bearing stars can appear below the CMD thresholds, I allowed
the selection of stars below those thresholds if they exhibited
mid-IR excess emission in data from WISE. This selection process has produced
22 candidates. All but one of the candidates are beyond the eastern boundary
of the field considered in my previous surveys of Taurus.
As discussed in Section~\ref{sec:spec}, 13 of the candidates have
spectroscopy available, all of which are confirmed to be young and have
been adopted as members. The remaining nine candidates that lack spectra
are presented in Table~\ref{tab:cand}, which includes astrometry
and photometry from Gaia DR3, radial and $UVW$ velocities, IR photometry
from 2MASS and WISE, disk classifications, and Taurus group assignments.

\section{A Search for Members of Young Associations near Taurus}
\label{sec:groupsearch}

\subsection{Identification of Candidate Members}
\label{sec:ident}

Many studies have found young stars in the general vicinity of
the Taurus clouds that have kinematics or ages that seem inconsistent
with membership in the groups associated with the clouds
\citep{bla56,wal88,neu95,wic96,luh06tau1,sle06,dae15,kra17,esp17}.
Some of those stars have been identified as possible members of small groups 
\citep{mam07,mam16,bel17,liu21,ker21,kro21} and larger associations 
\citep{luh18,liu20,gag20}.
I have attempted to perform a systematic search for young associations
that are near Taurus spatially and kinematically using Gaia DR3.
I have considered Gaia sources with distances between 80 and
250~pc, right ascensions between 35$\arcdeg$ and 90$\arcdeg$, declinations
between $-10\arcdeg$ and 35$\arcdeg$, $\sigma_{\pi}<0.5$~mas, and proper 
motion offsets of $\mid\Delta\mu_{\alpha,\delta}\mid<15$~mas~yr$^{-1}$ 
when calculated for the median velocity of Taurus.
For reference, the Pleiades cluster is within that volume of space
but is beyond the thresholds for proper motion offsets and the opposite is
true for the $\alpha$~Per cluster.
I selected candidate young low-mass stars with colors of
$G_{\rm BP}-G_{\rm RP}>1$ and positions in CMDs 
above the sequence of Pleiades members \citep{sta07},
which has an age of $\sim120$~Myr \citep{sta98,dah15}.
Among those candidates, I identified associations based on their clustering
in proper motion offsets and spatial positions. For each association that
was found, sequences of members were apparent in the two CMDs, so I defined
boundaries that followed the lower envelopes of the sequences and used them
to further refine the sample of candidate members. I defined criteria
for the proper motion offsets that captured the refined sample of
low-mass stellar candidates and I used those criteria and the CMD
boundaries to select candidate members across the full range of stellar masses.
For each association, the spatial positions of the resulting kinematic
and CMD candidates exhibited a concentration surrounded by a very sparse
population extending to the boundaries of the survey volume. The latter
are likely dominated by field stars, so for a given association I report
only the candidates in the spatial volume where they are concentrated.
As done in Section~\ref{sec:tausearch} for Taurus, I have allowed the
selection of candidates that appear below the CMD thresholds if they exhibit 
mid-IR excess emission in data from WISE. I have not attempted to identify 
white dwarfs in the associations, which would fall below the CMD thresholds
as well.

After arriving at a sample of candidates for a given association, I searched
for companions to those candidates that are in Gaia DR3 but did not satisfy 
the selection criteria. I retrieved DR3 sources that have separations of
$\leq5\arcsec$ from the candidates and that 
(1) fail the kinematic criteria but satisfy the photometric criteria and
share roughly similar parallaxes and proper motions as their neighboring
candidates,
(2) lack measurements of proper motion and parallax but satisfy the
photometric criteria when adopting the distances of their neighboring
candidates, or
(3) satisfy the kinematic criteria but lack the photometry for the CMDs.
The resulting objects have been included in my samples of candidate
members of the associations.

In the next section, I describe the compilation of various data that
are available for the candidates, which includes spectral classifications 
and radial velocity measurements.  I have rejected each candidate that 
(1) has a nondetection of Li at 6707~\AA\ that would be inconsistent with 
membership in the association in question
or (2) has a radial velocity that produces a discrepant space velocity 
relative to the bulk of the candidates for a given association. 
In the latter scenario, a candidate is retained if its discrepant velocity 
may be due to binarity as indicated by the presence of multiple velocity
measurements that span a large range.

Eight associations are detected in the vicinity of Taurus through my analysis.
Their names and numbers of candidates are 32~Ori (169),
$\mu$~Tau (354), 93~Tau (190), 69~Ori (591), HD~284346 (78), HD~33413 (47),
HD~35187 (65), and V1362~Tau (53). 
The 32~Ori association was discovered by \citet{mam07} and surveyed
by \citet{bel17}. My catalog of candidates for 32~Ori from Gaia DR3 is
presented in \citet{luh22o}. Catalogs for the other seven associations 
are presented in this work. The $\mu$~Tau association was discovered 
with Gaia DR2 by \citet{liu20} and \citet{gag20}. 
93~Tau originated as group 29 from \citet{oh17}, which consisted of
nine stars selected from the first data release (DR1)
of Gaia and expanded to 91 candidates with Gaia DR2 \citep{luh18}.
The brightest star in my new sample from DR3 is 93~Tau, which I have
adopted as the name for that association. 
As with 93~Tau, I have named the five remaining associations after their
brightest members. In \citet{luh22o} and Section~\ref{sec:compare}, 
I compare my samples of candidates for the eight associations to those from
previous studies.

\begin{figure}
\epsscale{1.1}
\plotone{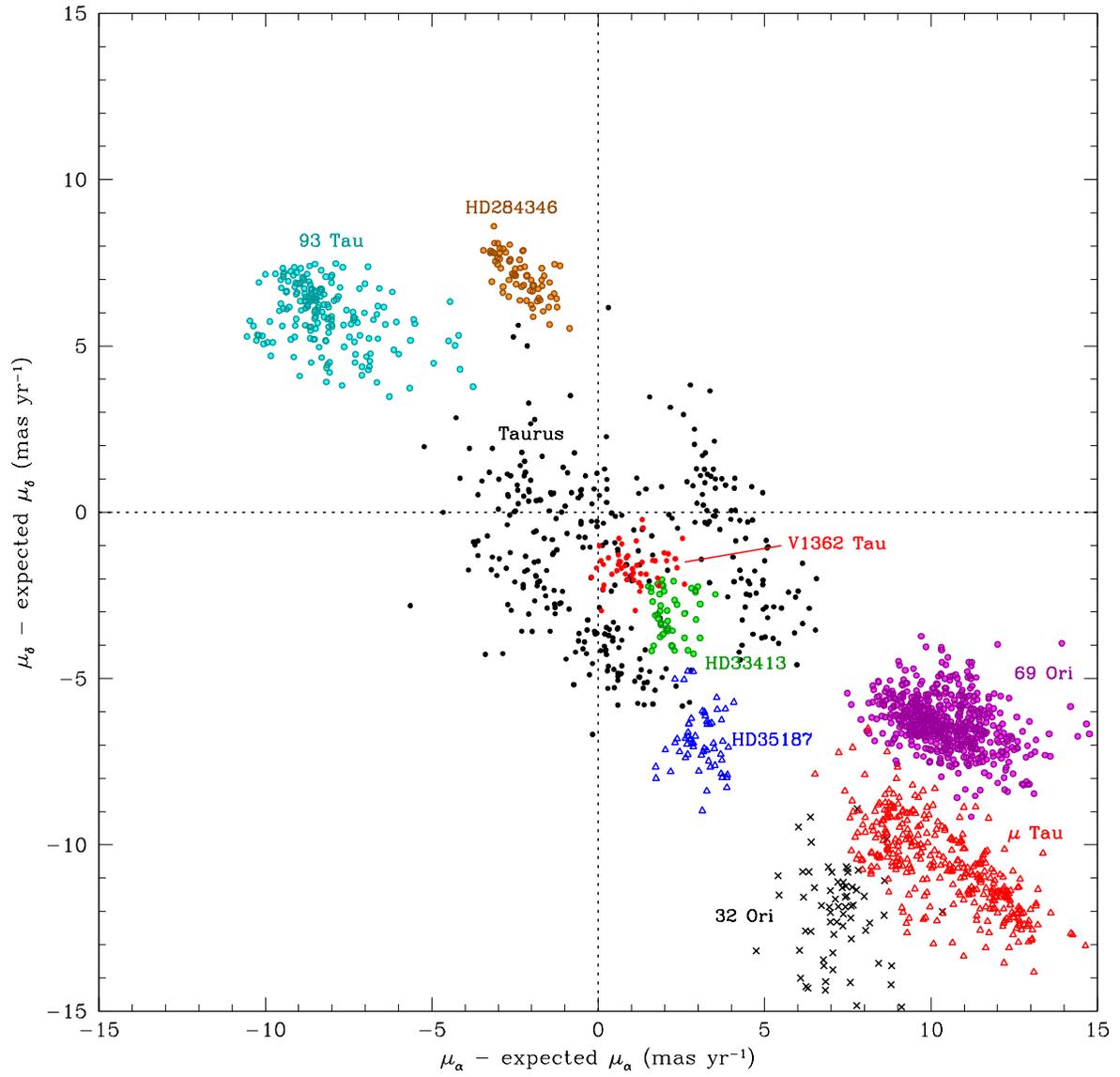}
\caption{
Proper motion offsets based on Gaia DR3 for members of Taurus 
(Table~\ref{tab:mem}) and candidate members of neighboring associations 
\citep[Table~\ref{tab:groups},][]{luh22o}.
The offsets are calculated for the median velocity of Taurus.}
\label{fig:pp5}
\end{figure}

In Figure~\ref{fig:pp5}, I have plotted the proper motion offsets
for the adopted members of Taurus and the candidate members of 
the eight neighboring associations, which are calculated for the
median velocity of Taurus. The diagonal alignment of the
associations is a reflection of the fact that their
velocities differ primarily in $V$ (Section~\ref{sec:uvw}).
Most of the associations have distinct kinematics in Figure~\ref{fig:pp5}
relative to Taurus. One of them, V1362~Tau, does overlap slightly
with a Taurus group, L1544. Their kinematics, spatial distributions, 
and ages are compared in Section~\ref{sec:ker21}.

In a diagram of proper motion offsets like the one in Figure~\ref{fig:pp5},
the distribution of offsets for a group of stars will be broader
for larger deviations from the velocity assumed when calculating
the offsets (i.e., when the offsets are larger), particularly if the group
is spread across a large area of sky.  For instance, the associations in
Figure~\ref{fig:pp5} that are farther from the origin like 93~Tau and 32~Ori
would be somewhat more tightly clustered (and near the origin) if their
offsets were calculated with their median velocities rather than that of Taurus.
As a result, the selection of candidates for an association is more refined
if performed with proper motion offsets calculated for the association's
velocity. Therefore, after identifying a given association via its clustering
in the proper motion offsets for the velocity of Taurus, I compiled
the available radial velocities for the initial sample of candidates,
calculated their space velocities, derived new proper motion offsets for
their median velocity, and repeated the identification of candidates via
the clustering of the new offsets.

\begin{figure}
\epsscale{1.2}
\plotone{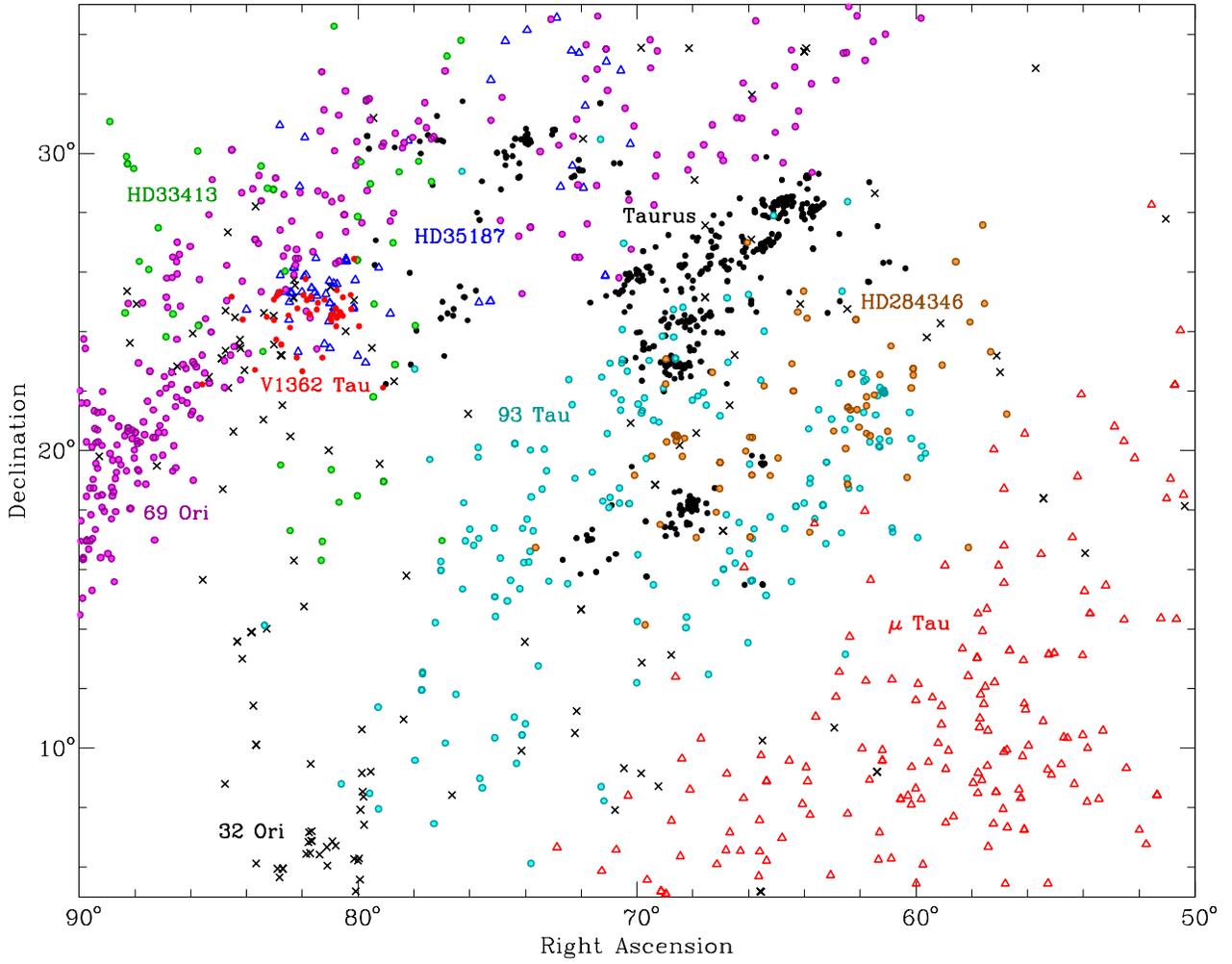}
\caption{
Equatorial coordinates for members of Taurus (Table~\ref{tab:mem}) and
candidate members of neighboring associations
\citep[Table~\ref{tab:groups},][]{luh22o}.}
\label{fig:map2}
\end{figure}

The distributions of the Taurus groups and the neighboring associations
on the sky are compared in a map of equatorial coordinates in
Figure~\ref{fig:map2}. 32~Ori, 93~Tau, and HD~284346 overlap with the main
complex of Taurus groups while four of the other associations are near
the two easternmost Taurus groups, L1517 and L1544.
The properties of the stellar populations in the associations near Taurus
(e.g., ages, spatial and kinematic distributions) are investigated in
Section~\ref{sec:pops}.

\subsection{Compilation of Data}
\label{sec:compilation2}

A compilation of data for the candidate members of 32~Ori is provided in
\citet{luh22o}. The candidates for the remaining seven associations
near Taurus are presented in Table~\ref{tab:groups}, which includes
source names from Gaia DR3 and previous studies; equatorial coordinates,
proper motion, parallax, RUWE, and photometry from Gaia DR3;
measurements of spectral types and the type adopted in this work;
distance estimate based on the Gaia DR3 parallax \citep{bai21}; 
the most accurate available radial velocity measurement that has an error 
less than 4~km~s$^{-1}$; {\tt VSCATTER} from APOGEE-2; $UVW$ velocities; the
designations and angular separations of the closest sources within $3\arcsec$ 
from 2MASS and WISE; flags indicating whether the Gaia source is the closest 
match in DR3 for the 2MASS and WISE sources; photometry from 2MASS and WISE; 
flags indicating whether excesses are detected in three WISE bands and a disk
classification if excess emission is detected (Section~\ref{sec:disks});
and the name of the association to which the candidate is assigned.
Some of the compiled spectral types have been measured in this work.
I have classified an IRTF/SpeX spectrum from \citet{zha18}, an optical
spectrum obtained during the observations with the MMT Red Channel Spectrograph 
\citep{sch89} described in \citet{esp19}, optical spectra of 51 candidates
in 93~Tau that I collected with the Gemini Multi-Object Spectrograph
\citep[GMOS;][]{hoo04} at the Gemini North telescope, and optical spectra
from LAMOST DR7 for 391 candidates.  Among the 1378 candidates in 
Table~\ref{tab:groups}, 556 have spectral classifications and 274 have 
measurements of radial velocities.

\section{Properties of the Stellar Populations in Taurus and Neighboring
Associations}
\label{sec:pops}

\subsection{Spectral Types and Extinctions}
\label{sec:spt}

Portions of my analysis of the stellar populations in Taurus and neighboring
associations require spectral types and extinctions for the members
that have parallax measurements.
Most of the adopted members of Taurus have spectral classifications 
(Table~\ref{tab:mem}) and extinction estimates \citep{esp19}, including
nearly all of those with parallax data. For the stars in Taurus and 
neighboring associations that have spectral types but no previous estimates of
extinction, I have derived extinctions from color excesses in 
$G_{\rm RP}-J$, $J-H$, $G-G_{\rm RP}$, or $G_{\rm BP}-G_{\rm RP}$ (in order of 
preference) relative to the intrinsic colors of young stars at a given spectral 
type \citep{luh22sc}. 

Spectroscopy is not available for the nine Taurus candidates 
(Table~\ref{tab:cand})\footnote{The Taurus candidates are not
projected against dark clouds, so their extinctions are expected to be low.
Eight of the nine candidates have colors that indicate extinctions of 
$A_K<0.05$.} and most of the candidates in the associations.
For those stars, I have estimated spectral types and extinctions by
dereddening their observed colors to the sequences of intrinsic colors
of young stars in diagrams of $G_{\rm BP}-G_{\rm RP}$ and
$G_{\rm RP}-J$ versus $J-H$ using the extinction curve from \citet{sch16}.
Stars with types of K5--M5 have the most accurate estimates of spectral
types and extinctions from that process since the reddening vectors are closest 
to perpendicular to the sequences of intrinsic colors among those types.
Because the Gaia filters are broad, the reddening relation for a color
that includes a Gaia band depends significantly on the intrinsic spectrum
and extinction of a star. I derived extinction relations among the bands
in question as a function of extinction and effective temperature in
the manner done by \citet{luh20u} for Gaia DR2 but with the filter profiles
for DR3 \citep{rie21}. For each star, I dereddened the observed colors
to the intrinsic colors of young stars using the reddening relations
for $A_K=0.1$ and a temperature that corresponds to the spectral type
implied by the observed color, producing an initial estimate of extinction.
The dereddening process was then iteratively repeated using
the reddening relations for the new extinction and temperature
(as implied by the spectral type) until converging on the final values.

\begin{figure}
\epsscale{1.1}
\plotone{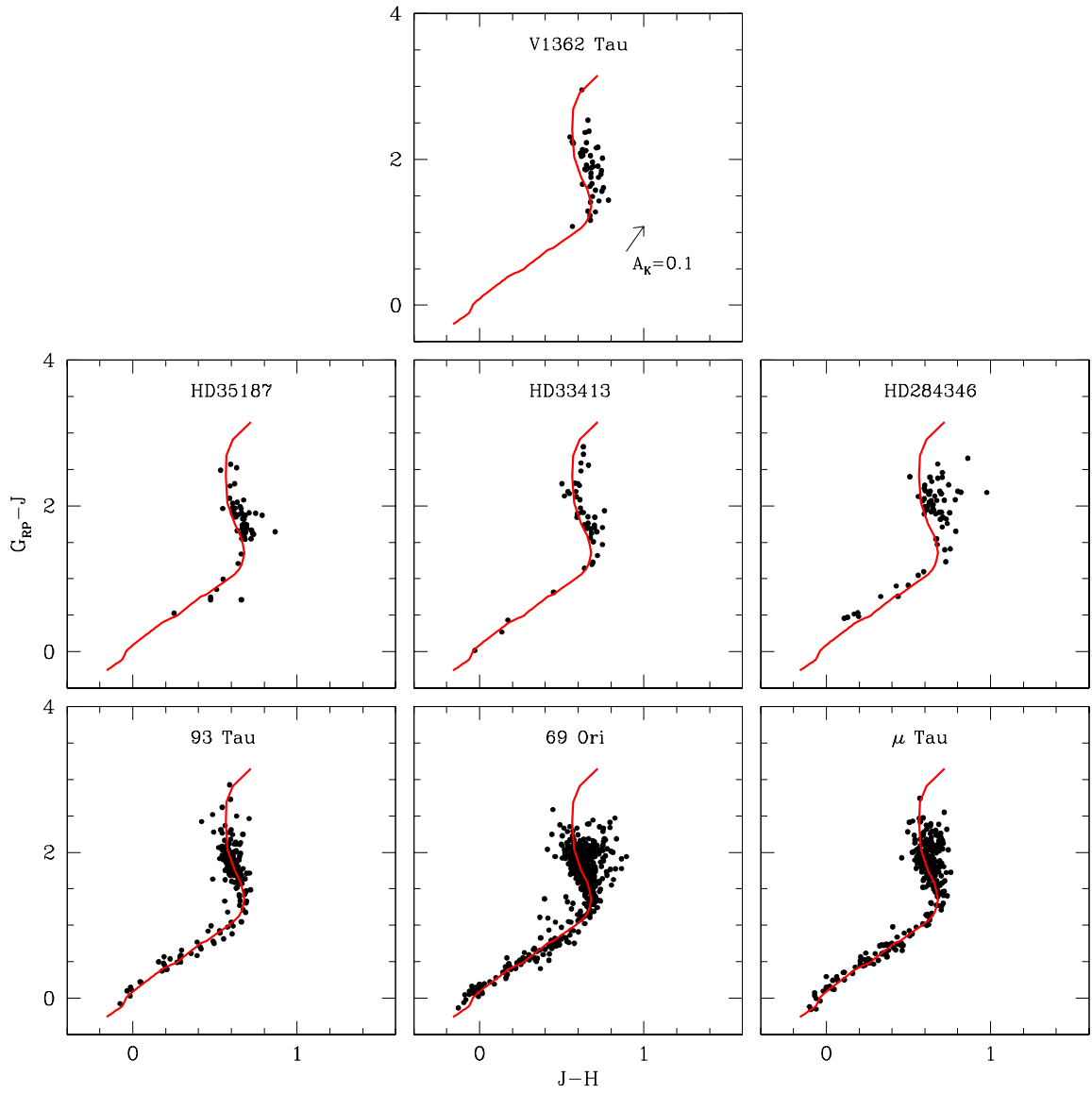}
\caption{
$G_{\rm RP}-J$ versus $J-H$ for candidate members of associations near Taurus. 
The intrinsic colors of young stars from B0--M9 are indicated 
\citep[red lines,][]{luh22sc}.}
\label{fig:cc}
\end{figure}

To illustrate the levels of extinction for the associations near Taurus,
I have plotted their candidate members with the typical intrinsic colors of
young stars in diagrams of $G_{\rm RP}-J$ versus $J-H$ in Figure~\ref{fig:cc}.
Among K5--M5 stars, which have the most accurate extinction estimates when
spectral types are not available, 
the median values of $A_K$ are 0.043 (V1362~Tau), 0.029 (HD~35187), 
0.021 (HD~33413), 0.055 (HD~284346), 0.003 (93~Tau), 0.012 (69~Ori),
and 0.015 ($\mu$~Tau). A similar diagram for 32~Ori indicates negligible
extinction \citep{luh22o}. As expected, extinction is lowest in the closest
associations. HD~284346 has the highest reddening since it is behind the
Taurus clouds. The western edge of 69~Ori also extends behind Taurus, resulting
in a small number of members at somewhat higher reddenings.

\subsection{Initial Mass Functions}

The census of Taurus from \citet{esp19} was estimated to have 
a high level of completeness for (1) spectral types earlier than M6--M7
at $A_J<1$ within a field encompassing all of the Taurus clouds
and (2) spectral types of $\lesssim$L0 at $A_J<1.5$ within fields
covering a large fraction of the known members. The completeness estimates
for those two samples were primarily based on Gaia DR2 and deep optical
and IR imaging, respectively. Since my new work with Gaia DR3 has produced
only minor changes to the census of Taurus, it is unnecessary to update
the analysis of the initial mass function (IMF) in Taurus from \citet{esp19}.

\begin{figure}
\epsscale{1.2}
\plotone{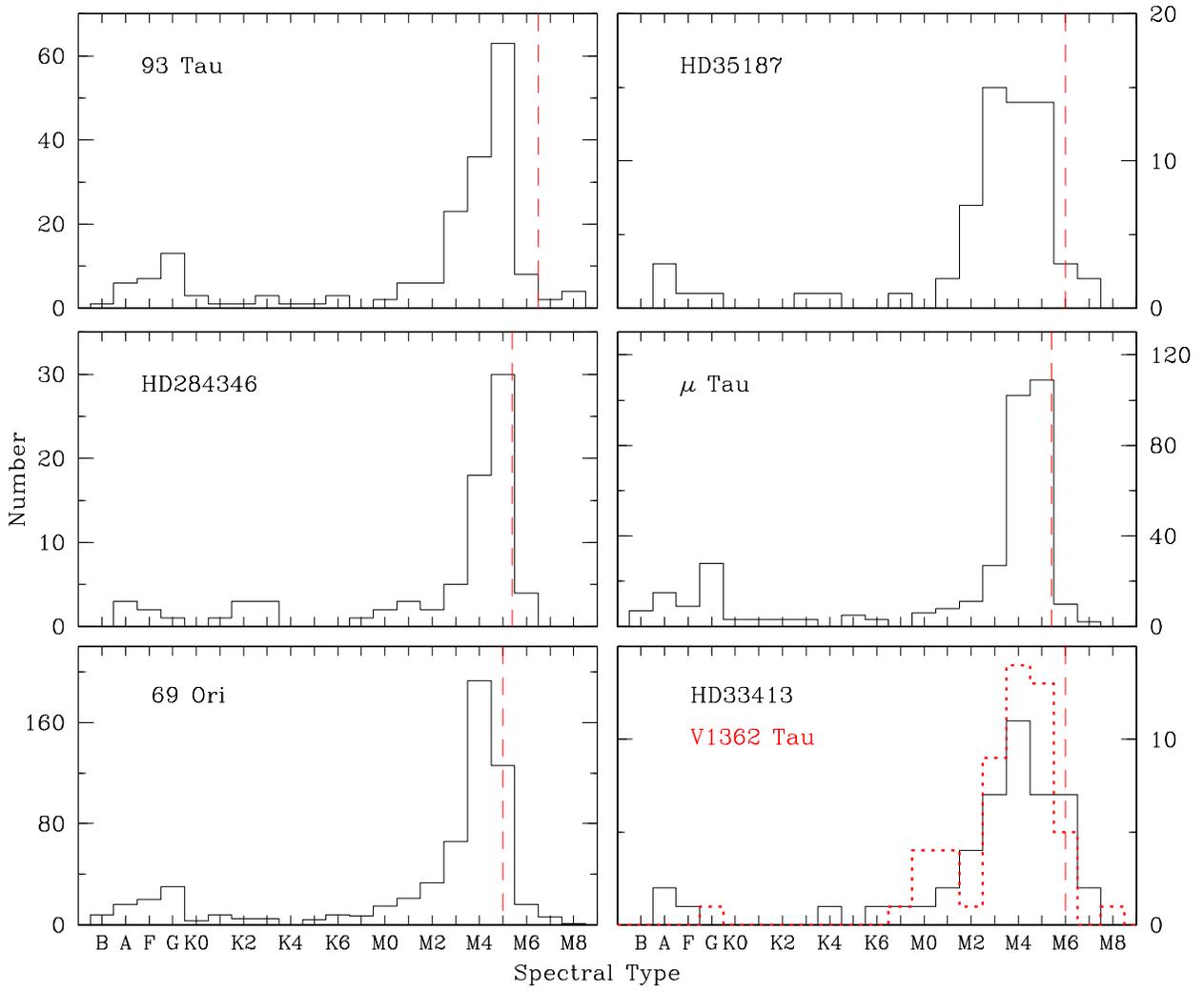}
\caption{
Histograms of spectral types for candidate members of associations near Taurus.
For stars that lack spectroscopy, spectral types have been estimated from
photometry (Section~\ref{sec:spt}). Completeness limits are indicated
(dashed lines).}
\label{fig:histo}
\end{figure}

For the associations near Taurus, I choose to use histograms of spectral types
as observational proxies for their IMFs, as done in my previous work in
Taurus and other nearby young populations.
The photometric estimates of spectral types from the previous
section are used for stars that lack spectroscopy.
\citet{luh22o} presented a histogram of spectral types for 32~Ori.
The histograms for the remaining associations near Taurus are 
plotted in Figure~\ref{fig:histo}. The completeness limits are marked,
which were derived in the manner described by \citet{luh22o}.
All of the associations exhibit a prominent peak near M5 
($\sim0.15$~$M_\odot$), resembling the distributions in Taurus \citep{esp19} 
and other nearby associations and star-forming regions \citep[e.g.,][]{luh22sc}.

\subsection{Ages}
\label{sec:ages}

The ages of young stellar populations can be estimated via their sequences
of low-mass stars in the Hertzsprung-Russell (H-R) diagram \citep{her15}.
For the Taurus groups and the neighboring associations, I have constructed
H-R diagrams in parameters that should minimize the combined errors associated
with extinction correction, disk-related emission, and the measurement
of photometry given the data that are available.
For the H-R diagrams in Taurus, I have selected spectral types as proxies
for effective temperatures because they are available for nearly all of the
Gaia-detected members and are not affected by extinction or disk emission
when measured properly. Because some Taurus members have significant extinction,
I have selected absolute magnitudes in $K_s$ from 2MASS ($M_K$) to represent
the luminosities, which is long enough in wavelength that extinction
is low while short enough in wavelength that emission from non-full disks
is negligible. The lower precision of the 2MASS photometry relative to
the Gaia data should be more than compensated by the smaller errors in the
extinction corrections.
For the associations near Taurus, the optimal approach is to construct
the H-R diagrams with $G_{\rm BP}-G_{\rm RP}$ and $M_{G_{\rm RP}}$
since these bands have high precision, are available for most of
the candidates, and are not subject to large errors in extinction
corrections given the low extinctions of the associations.
Since extinction varies significantly among the members of Taurus,
their individual extinction estimates are used to correct their
photometry in $K_s$ (Section~\ref{sec:spt}).
The extinctions within the other associations span very small ranges
(Figure~\ref{fig:cc}), so I have corrected the Gaia photometry using
the median value of $A_K$ in a given association (Section~\ref{sec:spt}).

In addition to the groups and associations surveyed in this work,
I have included in my age analysis two young clusters in the vicinity
of Taurus that are rich and well-studied: the Pleiades and $\alpha$~Per.
For the Pleiades, I have adopted the sample of members from \citet{sta07}
after rejecting a few kinematic and photometric outliers based on Gaia DR3.
For $\alpha$~Per, I have identified a sample of candidate members within
a $3\arcdeg$ radius field containing the cluster core using
data from Gaia DR3 and the methods from Section~\ref{sec:groupsearch}.
Alternative catalogs of members of these two clusters selected with Gaia DR2
are available from \citet{lod19}.
The Gaia data for the Pleiades have been corrected for an extinction
of $A_K=0.012$ \citep{sta07}. An extinction correction of $A_K=0.014$
has been applied to the candidates in $\alpha$~Per, which is the median value
derived from color-color diagrams in the manner described earlier
for other associations. 

\begin{figure}
\epsscale{1.1}
\plotone{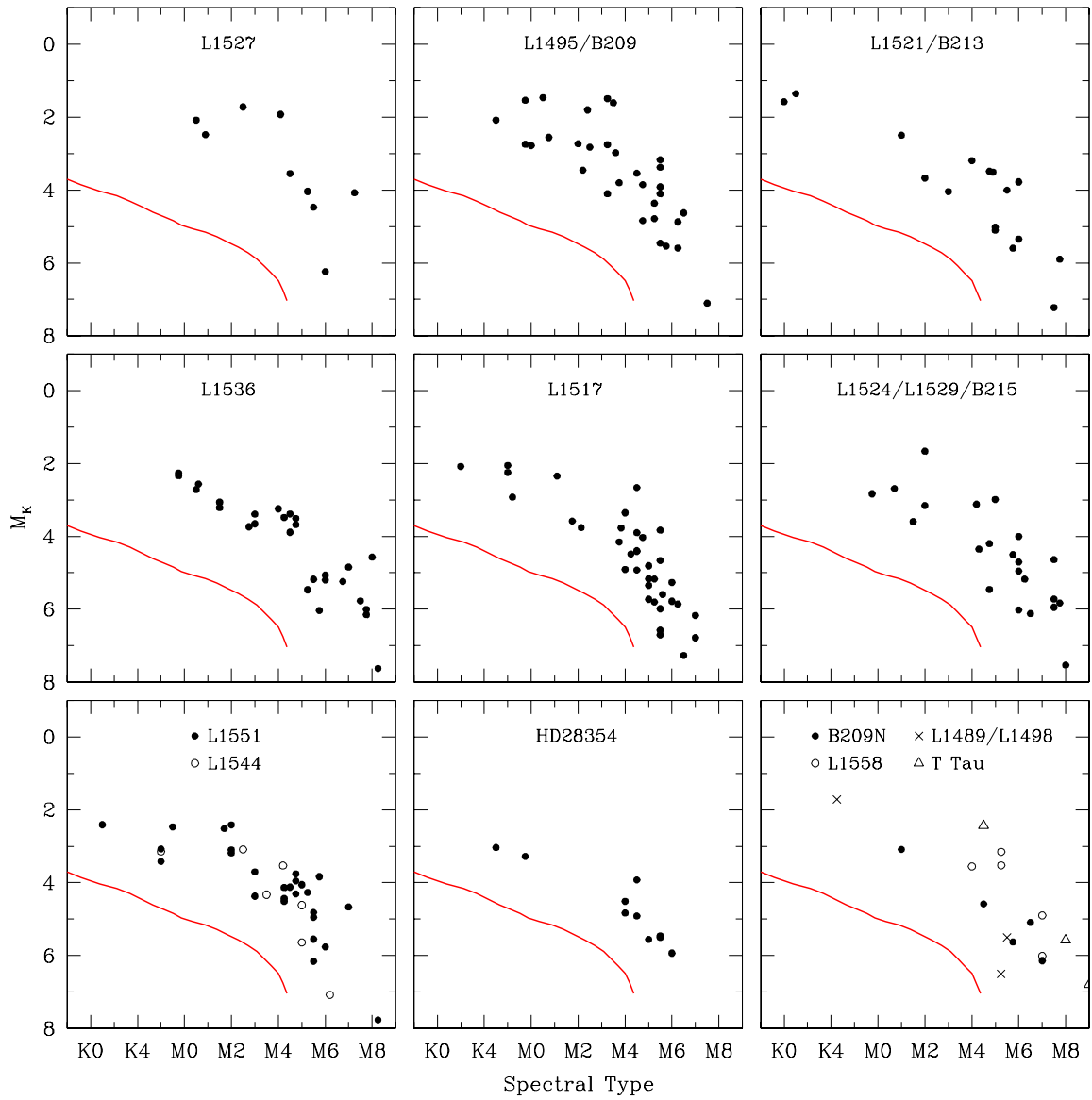}
\caption{
Extinction-corrected $M_K$ versus spectral type for the Taurus groups.
A fit to the single-star sequence of the Pleiades is indicated (red solid line).}
\label{fig:hr}
\end{figure}

\begin{figure}
\epsscale{1.1}
\plotone{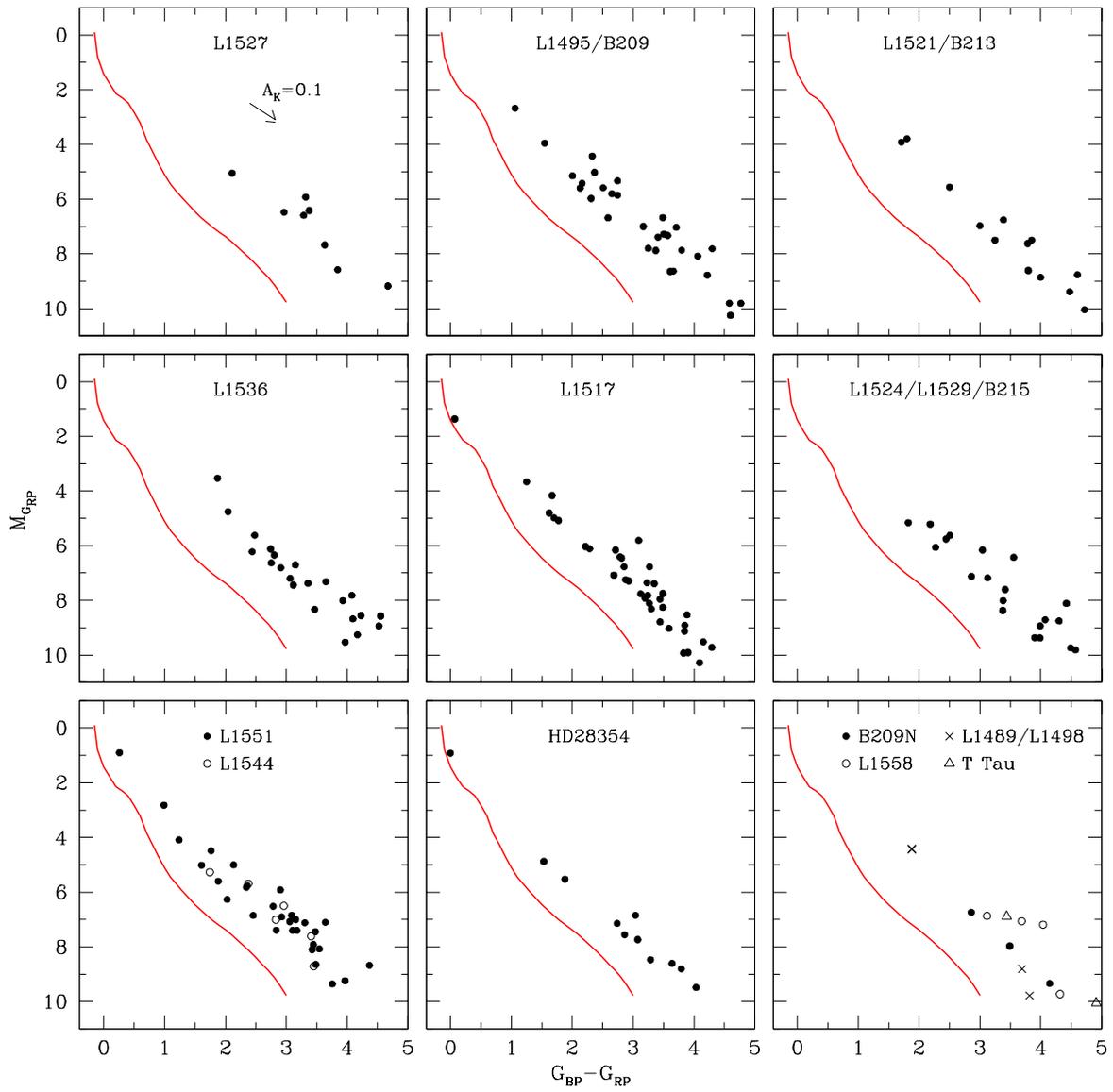}
\caption{
$M_{G_{\rm RP}}$ versus $G_{\rm BP}-G_{\rm RP}$ for the Taurus groups.
A fit to the single-star sequence of the Pleiades is indicated (red solid line).
}
\label{fig:cmd2}
\end{figure}

\begin{figure}
\epsscale{1.1}
\plotone{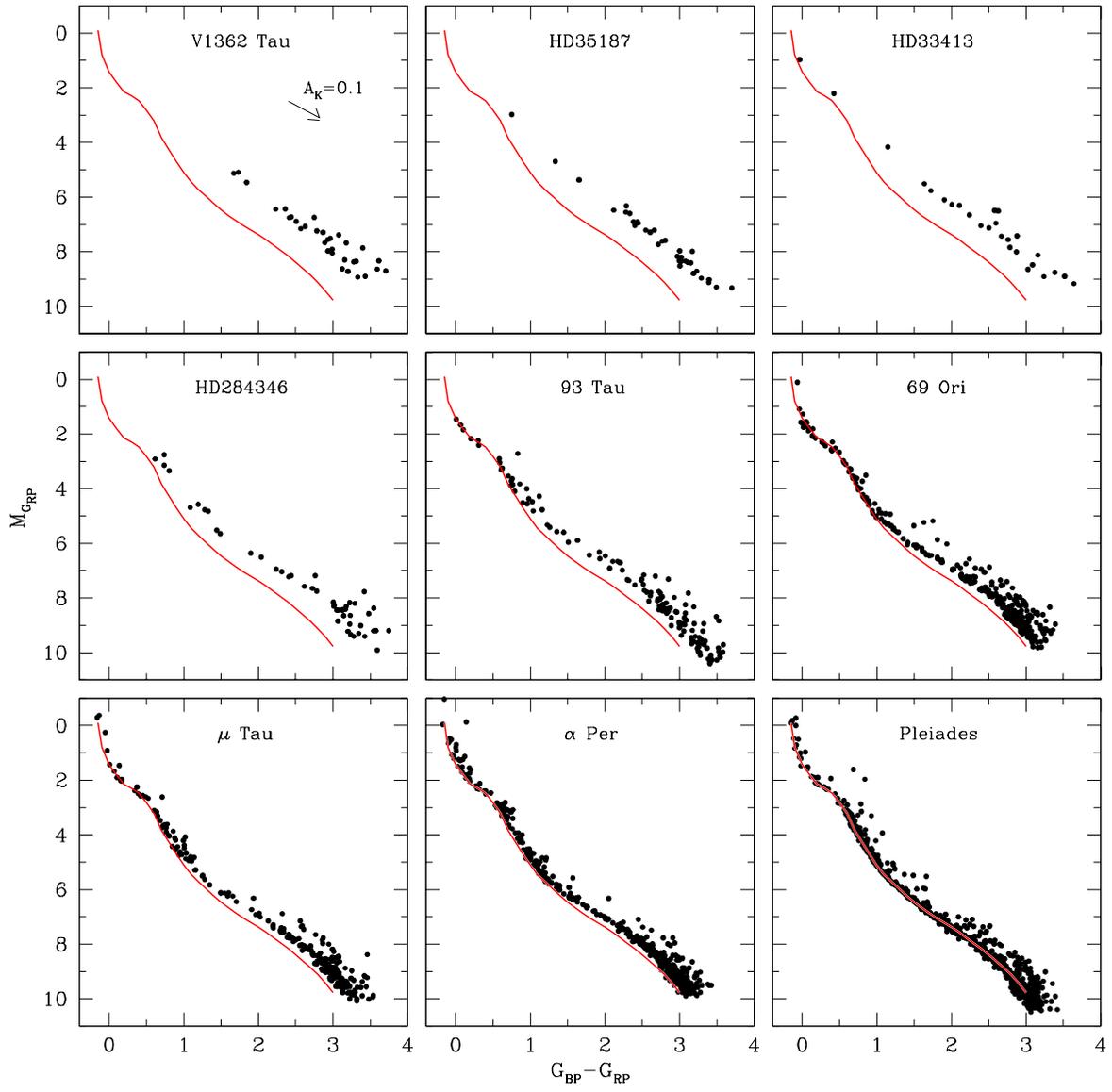}
\caption{
$M_{G_{\rm RP}}$ versus $G_{\rm BP}-G_{\rm RP}$ for associations near Taurus
and the $\alpha$~Per and Pleiades clusters.
These data have been corrected for the median extinction of each association. 
A fit to the single-star sequence of the Pleiades is indicated (red solid line).
}
\label{fig:cmd3}
\end{figure}

For all H-R diagrams, I have excluded stars with full disks to
mitigate contamination of the photometry by disk-related emission.
Because the uncertainties in the age estimates 
depend in part on the errors in the distances and photometry, I have included
only the candidates that have RUWE$<$1.6, $\sigma_{\pi}<0.1$~mas, 
$\sigma_{BP}<0.1$, and $\sigma_{RP}<0.1$ in the CMDs for the associations
near Taurus. Meanwhile, Taurus members with $\sigma_{\pi}<1$~mas
are considered, but for those with $\sigma_{\pi}>0.1$~mas, the median distances
of their groups are adopted when calculating $M_K$.
For the CMD of 69~Ori, I have omitted the small subset of candidates that are
behind the Taurus clouds since they have higher extinctions.
The diagrams of extinction-corrected $M_K$ versus spectral type for the
Taurus groups are shown in Figure~\ref{fig:hr}.
In each diagram, I have marked a fit to the single-star sequence of the Pleiades
in a diagram of $M_{G_{\rm RP}}$ versus $G_{\rm BP}-G_{\rm RP}$, which has
been converted to $M_K$ and spectral type
using the typical intrinsic colors of young stars \citep{luh22sc}.
Although they are not used in the age analysis, I also have plotted the
observed CMDs (no extinction corrections) of the Taurus groups in
Figure~\ref{fig:cmd2} for reference.
In Figure~\ref{fig:cmd3}, I present the extinction-corrected CMDs for the 
associations near Taurus and the $\alpha$~Per and Pleiades clusters.
A CMD in the same bands for 32~Ori is provided in \citet{luh22o}.
In Figures~\ref{fig:hr}--\ref{fig:cmd3}, the populations are shown in order of 
the ages implied by their sequences with the exception of one H-R diagram that 
combines the four smallest Taurus groups (B209N, L1558, L1489/L1498, T~Tau).
As found in many previous studies, the sequences of low-mass stars are
broadest in absolute magnitude at the youngest ages and become narrower at
older ages. At old enough ages, the sequences are sufficiently well-defined
that separate sequences of single stars and unresolved binaries are detected,
as shown in Figure~\ref{fig:cmd3}.

To estimate ages from the sequences of low-mass stars in the H-R diagrams,
I have considered stars within a range of $G_{\rm BP}-G_{\rm RP}$
across which the sequences are predicted to fade at a similar rate 
for ages of 1--80~Myr (i.e., the isochrones maintain a similar shape over time).
Based on theoretical evolutionary models \citep{bar15,cho16,dot16,fei16},
I have selected $G_{\rm BP}-G_{\rm RP}=1.4$--2.8 for the associations, which
corresponds to spectral types of $\sim$K4--M4 \citep{luh22sc}, temperatures of
$\sim$3300--4400~K, and masses of $\sim$0.2--1~$M_\odot$.
I have closely compared the median sequences within that color range for 
32~Ori, Upper Centaurus-Lupus/Lower Centaurus-Crux (UCL/LCC), and the 
associations and clusters in Figure~\ref{fig:cmd2},
finding that they do not show significant changes in shape with age 
until reaching the oldest population, the Pleiades, whose the sequence
becomes slightly steeper at $G_{\rm BP}-G_{\rm RP}>2.3$ in agreement
with model predictions. To alleviate the small number statistics in the 
Taurus groups, I extend the range of spectral types considered in their
age analysis to M5.

\begin{figure}
\epsscale{1.1}
\plotone{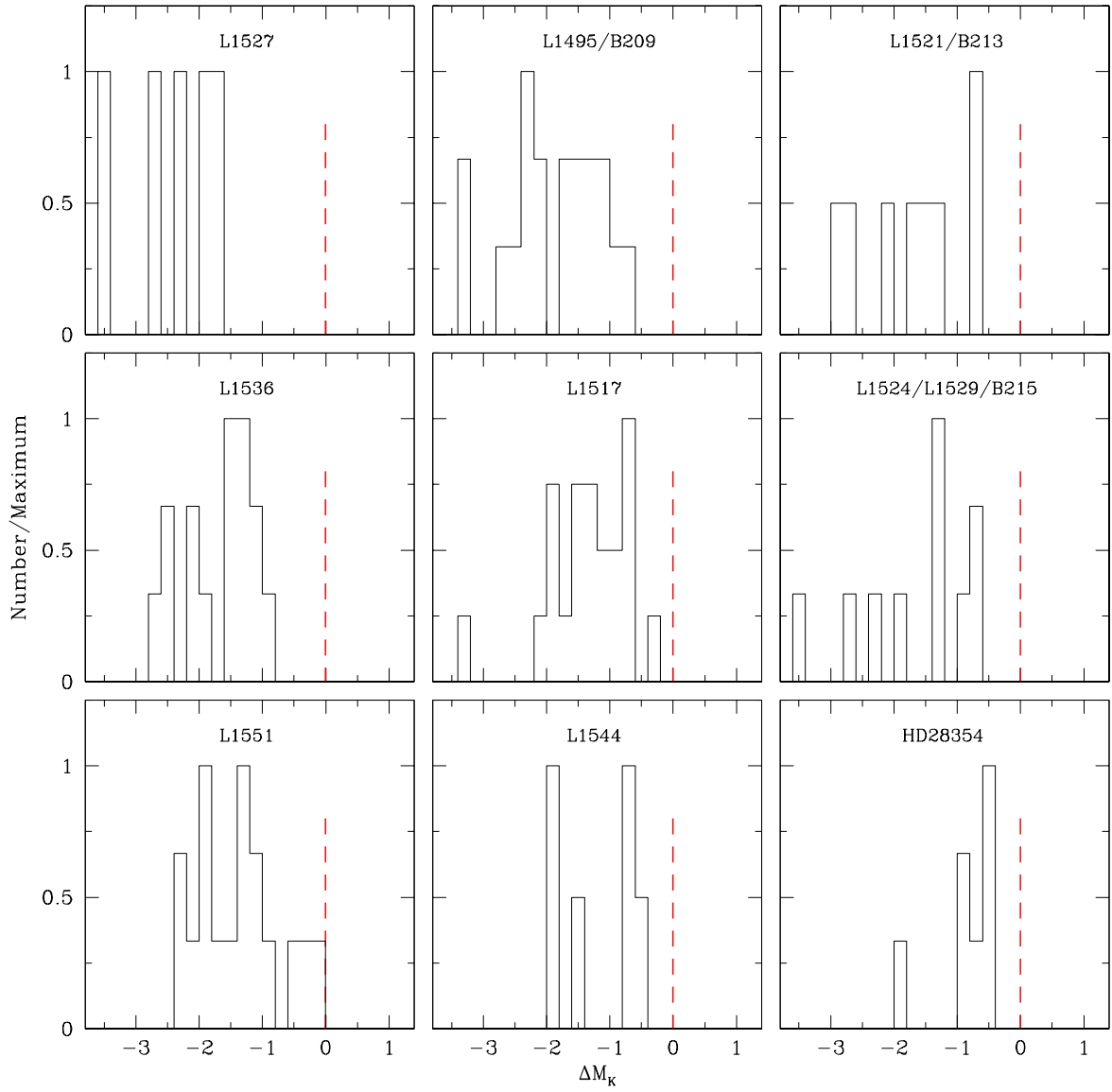}
\caption{
Histograms of offsets in $M_K$ from the median CMD sequence for UCL/LCC for
members of the Taurus groups with spectral types of 
K4--M5 (Figure~\ref{fig:hr}). Negative values correspond to brighter 
magnitudes and younger ages.}
\label{fig:at}
\end{figure}

\begin{figure}
\epsscale{1.1}
\plotone{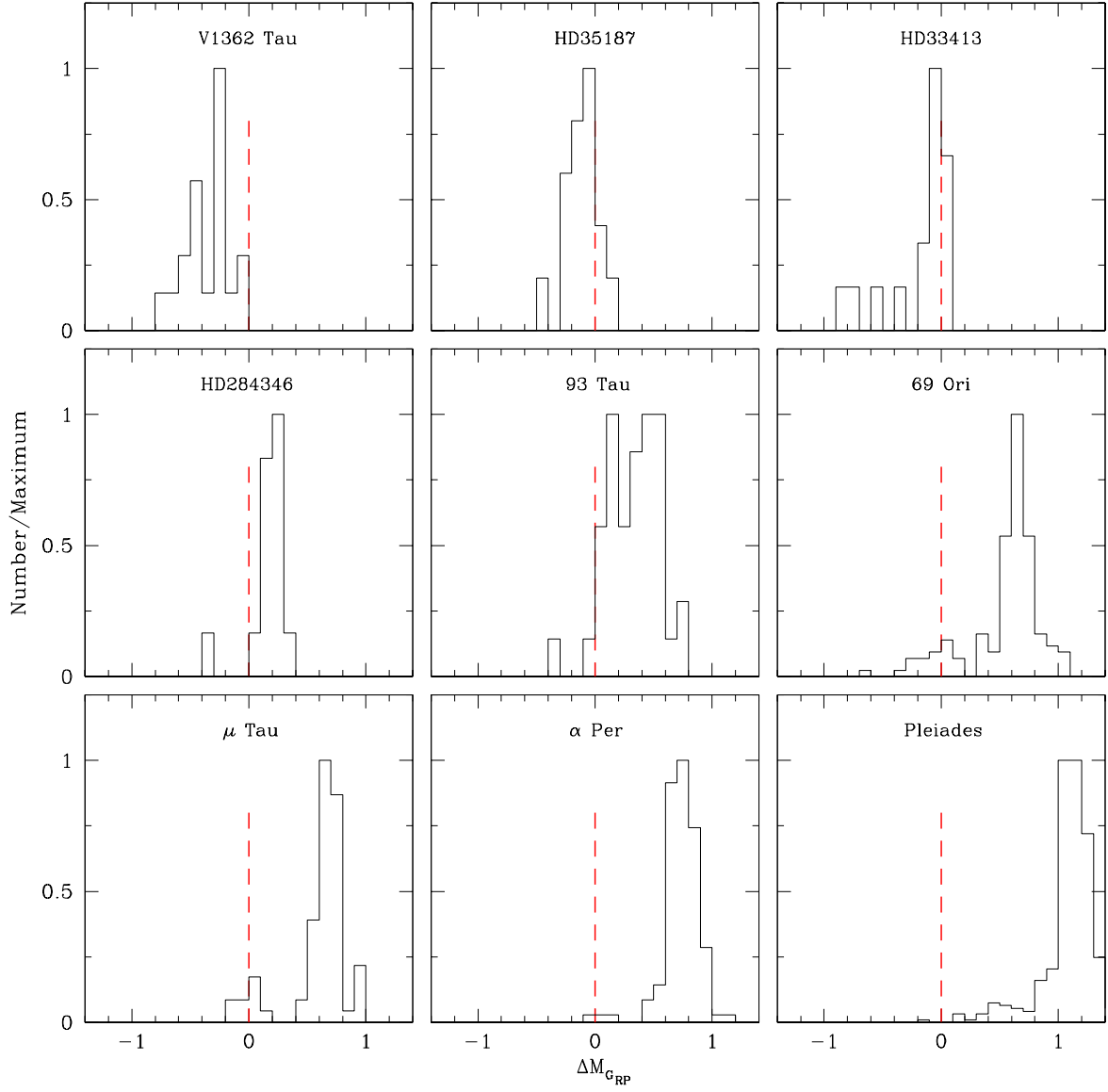}
\caption{
Histograms of offsets in $M_{G_{\rm RP}}$ from the median CMD sequence for
UCL/LCC for low-mass stars in associations near Taurus and the 
$\alpha$~Per and Pleiades clusters (Figure~\ref{fig:cmd3}).
Negative values correspond to brighter magnitudes and younger ages.}
\label{fig:ag}
\end{figure}

For the K4--M5 Taurus members in Figure~\ref{fig:hr} and
the stars between $G_{\rm BP}-G_{\rm RP}=1.4$--2.8 in Figure~\ref{fig:cmd3},
I have calculated offsets in $M_K$ and $M_{G_{\rm RP}}$ from a fit to the median
of the sequence for UCL/LCC \citep{luh22sc}, which is provided in 
Table~\ref{tab:fit}. That median sequence was 
derived in the CMD and then converted to spectral type and $M_K$ using the
intrinsic colors of young stars as a function of spectral types
from \citet{luh22sc}. Histograms of the resulting offsets in $M_K$
and $M_{G_{\rm RP}}$ are presented in Figures~\ref{fig:at} and \ref{fig:ag},
respectively. Similar measurements for 32~Ori are found in \citet{luh22o}.
I have not attempted to estimate the ages of the 
four smallest Taurus groups, so their histograms are not shown.
For the Pleiades, the offsets for $G_{\rm BP}-G_{\rm RP}<2.3$
are excluded since the cluster is old enough that those stars are no longer
fading at the same rate as younger stars.
I have calculated the median and the median absolute deviation (MAD) of the 
offsets in each population, which appear as $\Delta$M in Table~\ref{tab:ages}.
I have derived ages from those median offsets by assuming that UCL/LCC
($\Delta$M=0) has an age of 20~Myr \citep{luh22sc} and that 
$\Delta$log~L/$\Delta$log~age$=-0.6$, which describes the typical 
evolution of isochrones that has been predicted for the temperature range
in question \citep{bar15,cho16,dot16,fei16}.
The resulting ages are included in Table~\ref{tab:ages}.
The quoted errors reflect only the MAD of the offsets and do not
include systematic errors, which may vary with age.
In \citet{luh22o}, I found that 32~Ori is coeval with UCL/LCC based on
this kind of analysis.

Eight of the nine Taurus groups in Table~\ref{tab:ages} have median ages
between $\sim$1 and 3~Myr while the remaining group, HD~28354, as an age
of $\sim$6~Myr. The age uncertainties are large enough that most of the
groups could be coeval. The ages of the associations near Taurus
that I have surveyed range from 13--56~Myr.
These ages are based on an adopted age of 20~Myr for UCL/LCC, which
in turn is tied to the lithium depletion boundary (LDB) age for the
$\beta$~Pic moving group \citep{bin16}, as discussed in \citet{luh22sc}.
In addition, my age of 124$\pm$24~Myr for the Pleiades is consistent
with estimates of $125\pm8$~Myr and $112\pm5$~Myr based on the LDB
\citep{sta98,dah15}.  Meanwhile, my age of $61\pm11$~Myr for $\alpha$~Per is
younger than the LDB age of $85\pm10$~Myr from \citet{bar04}.
An age younger than the LDB value also has been produced for $\alpha$~Per by 
isochrone fitting of a CMD from Gaia DR2 \citep[70~Myr,][]{bab18}.

\subsection{Circumstellar Disks}
\label{sec:disks}

As mentioned in Section~\ref{sec:compilation}, disk classifications for my
adopted members of Taurus are provided in \citet{esp19} and 
Table~\ref{tab:spec}.  For the candidate members of the associations near 
Taurus, I have used mid-IR photometry from WISE to search for evidence of disks.
The WISE images were obtained in bands centered at 3.4, 4.6, 12, and 22~$\mu$m,
which are denoted as W1, W2, W3, and W4, respectively.
The 1378 candidates in the associations have 1291 matching sources from
WISE. If a close pair of candidates has the same WISE source as their closest
match, the WISE designation appears in both of their entries in 
Table~\ref{tab:groups}, but the disk measurements are listed
only for the candidate that is closest to the WISE source.
I have visually inspected the AllWISE Atlas images of the WISE sources
to check for detections that are false or unreliable, which are
marked by a flag in Table~\ref{tab:groups}.

\begin{figure}
\epsscale{1.2}
\plotone{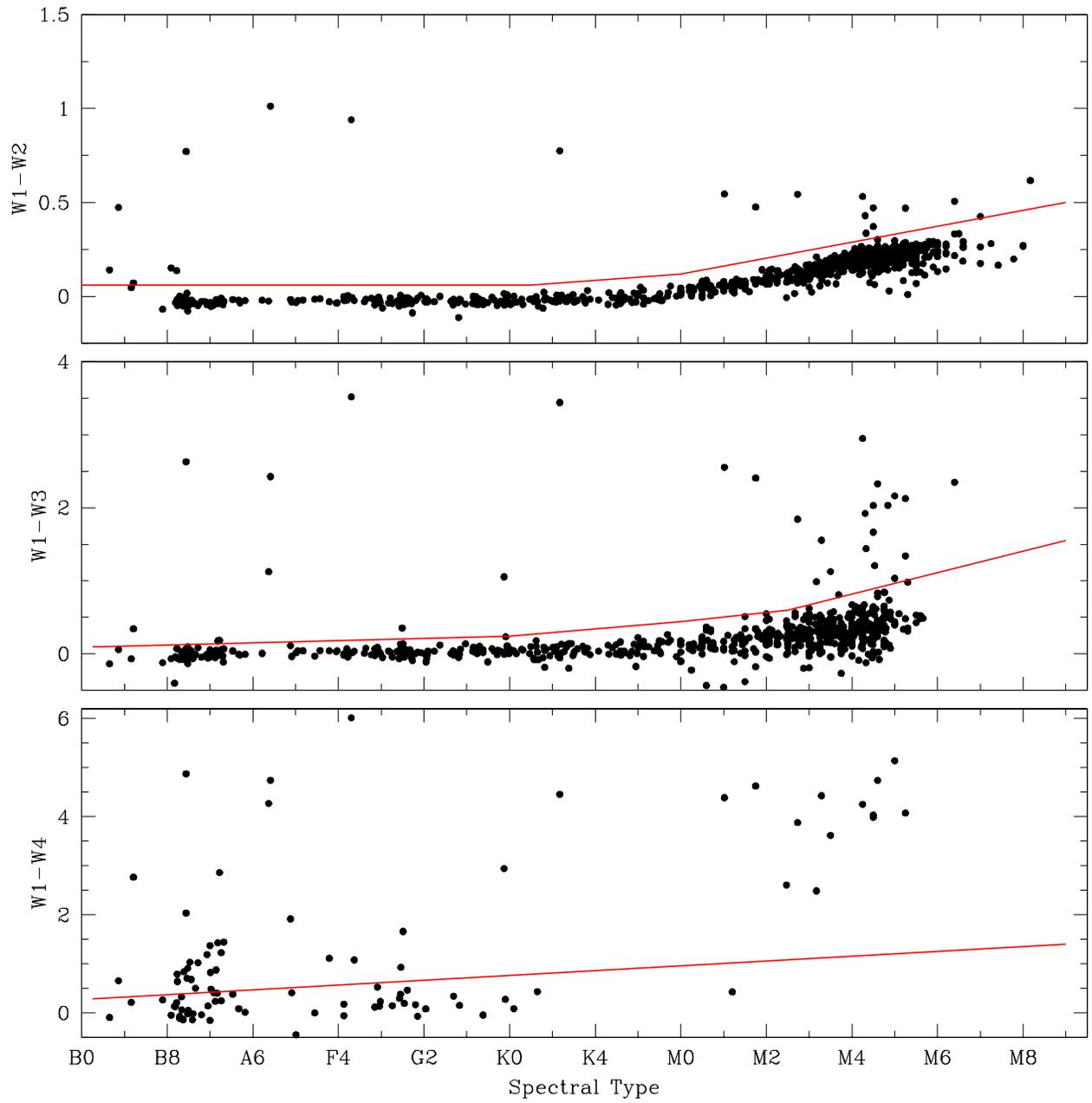}
\caption{
IR colors versus spectral type for candidate members of associations
near Taurus. For stars that lack spectroscopy, spectral types have been
estimated from photometry.  In each diagram, the tight sequence of blue 
colors corresponds to stellar photospheres. The thresholds used for identifying 
color excesses from disks are indicated \citep[red solid lines,][]{luh22disks}.
}
\label{fig:exc1}
\end{figure}

I have used W1$-$W2, W1$-$W3, and W1$-$W4 to detect excess emission from disks
among the 1291 WISE sources \citep{luh22sc}. 
Those colors are plotted versus spectral type in Figure~\ref{fig:exc1}.
The W2 data at W2$<$6 have been omitted since
they are subject to significant systematic errors \citep{cut12b}.
Photometric estimates of spectral types are adopted for stars that lack
spectroscopy (Section~\ref{sec:spt}).
In each of the three colors in Figure~\ref{fig:exc1}, most stars form
a well-defined sequence that corresponds to stellar photospheres.
A small number of stars have redder colors that indicate the presence of
IR excess emission. In Figure~\ref{fig:exc1}, I have marked the threshold for 
each color that was used by \citet{luh22disks} for identifying color excesses.
If a star appears above a threshold but a detection at a longer wavelength
is consistent with a photosphere, an excess is not assigned to the first band.
Table~\ref{tab:groups} includes flags that indicate whether excesses are 
present in W2, W3, and W4. The IR excess for the A-type star HD~284470 
was flagged as confused by \citet{reb11}, likely because of blending with a 
fainter companion at a separation of $10\arcsec$. However, I find that the two 
stars are sufficiently resolved that the WISE photometry should be reliable.

\begin{figure}
\epsscale{1.4}
\plotone{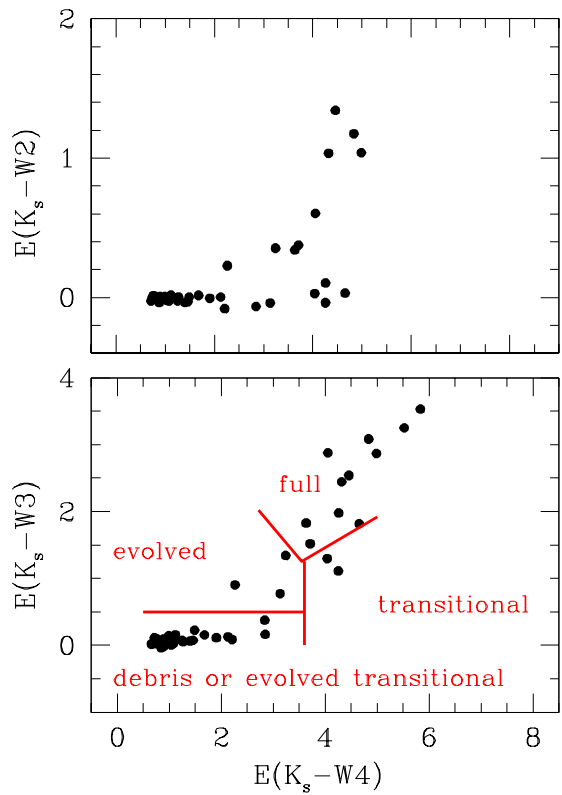}
\caption{
IR color excesses for candidate members of associations near Taurus.
The boundaries used for assigning disk classes are shown in the bottom
diagram (red solid lines).
}
\label{fig:exc2}
\end{figure}

I have classified the evolutionary stages of the detected disks from among
the following options: full disk, transitional disk, evolved disk, evolved
transitional disk, and debris disk \citep{ken05,rie05,her07,luh10,esp12}.
All of these classes except for the last one are primordial disks.
The classes are assigned based on the sizes of the excesses in
$K_s-$W3 and $K_s-$W4 \citep{luh12,esp14,esp18}.
I have calculated the color excesses, E($K_s-$W3) and E($K_s-$W4), by
subtracting the expected photospheric color for a given spectral type
\citep{luh22sc}. The resulting excesses are presented in Figure~\ref{fig:exc2}
with the criteria for the disk classes \citep{esp18}. To illustrate the sizes
of the excesses in W2, I have included E($K_s-$W2) as well.
As shown in Figure~\ref{fig:exc2}, the same criteria apply to
debris and evolved transitional disks, which are indistinguishable in
mid-IR photometry. Sources that lack excesses in any of the WISE bands 
are omitted from Figure~\ref{fig:exc2} and are listed as class~III. 

IR excesses are detected for 51 of the WISE sources, 31 of which have had disks
reported in previous work 
\citep{wal88,oud92,the94,luh06tau2,reb11,mcd12,mcd17,esp14,cot16,the17,liu21}.
The IR excess sources are classified as 18 full, two transitional, three 
evolved, three evolved or transitional, and 25 debris or evolved transitional
disks.
Most of the disk-bearing stars have not been previously recognized as members
of associations. My age estimates for those associations can be useful for
interpreting observations of the disks. In addition, primordial disks are
rare at the ages in question \citep{luh22disks}, making them valuable
for studies of disk evolution \citep{bou16,sil16,sil20,mur18,fla19,lee20}.
Three of the full disks are in 
well-known Herbig Ae/Be systems (HD~35187, CQ~Tau, MWC~758), all of which are 
candidate members of the HD~35187 association (18~Myr, Section~\ref{sec:ages}).
All but one of the disks classified as debris or evolved transitional have
spectral types of $\leq$G0. For the stars with IR excesses that lack 
spectroscopy, measurements of spectral classifications and radial velocities 
would be useful for confirming their youth and better constraining their 
membership.

\subsection{Spatial and Kinematic Distributions}
\label{sec:uvw}

Several aspects of the spatial and kinematic distributions in
Taurus and its neighboring associations can be investigated using the
astrometry from Gaia DR3 and the radial velocities (and corresponding
$UVW$ velocities) that have been compiled in Sections~\ref{sec:compilation} 
and \ref{sec:compilation2} (Tables~\ref{tab:mem} and \ref{tab:groups}).
For each group and association, I have calculated the median values of the 
parallactic distances, proper motions, and $UVW$ velocities and the standard 
deviations of the velocities, excluding the stars flagged as astrometric 
outliers in Tables~\ref{tab:mem}. I also have omitted sources with 
{\tt VSCATTER}$>3$~km~s$^{-1}$ from the velocity calculations. 
The results for the Taurus groups and the seven associations surveyed
in this work are presented in Table~\ref{tab:uvw}. The median velocity 
for 32~Ori is $U, V, W = -12.9, -18.9, -8.9$~km~s$^{-1}$ \citep{luh22o}.
From my sample of candidate members of $\alpha$~Per, I have calculated
a median velocity of $U, V, W = -14.1, -23.5, -6.8$~km~s$^{-1}$, which
is similar to the value for $\mu$~Tau.

\begin{figure}
\epsscale{1.2}
\plotone{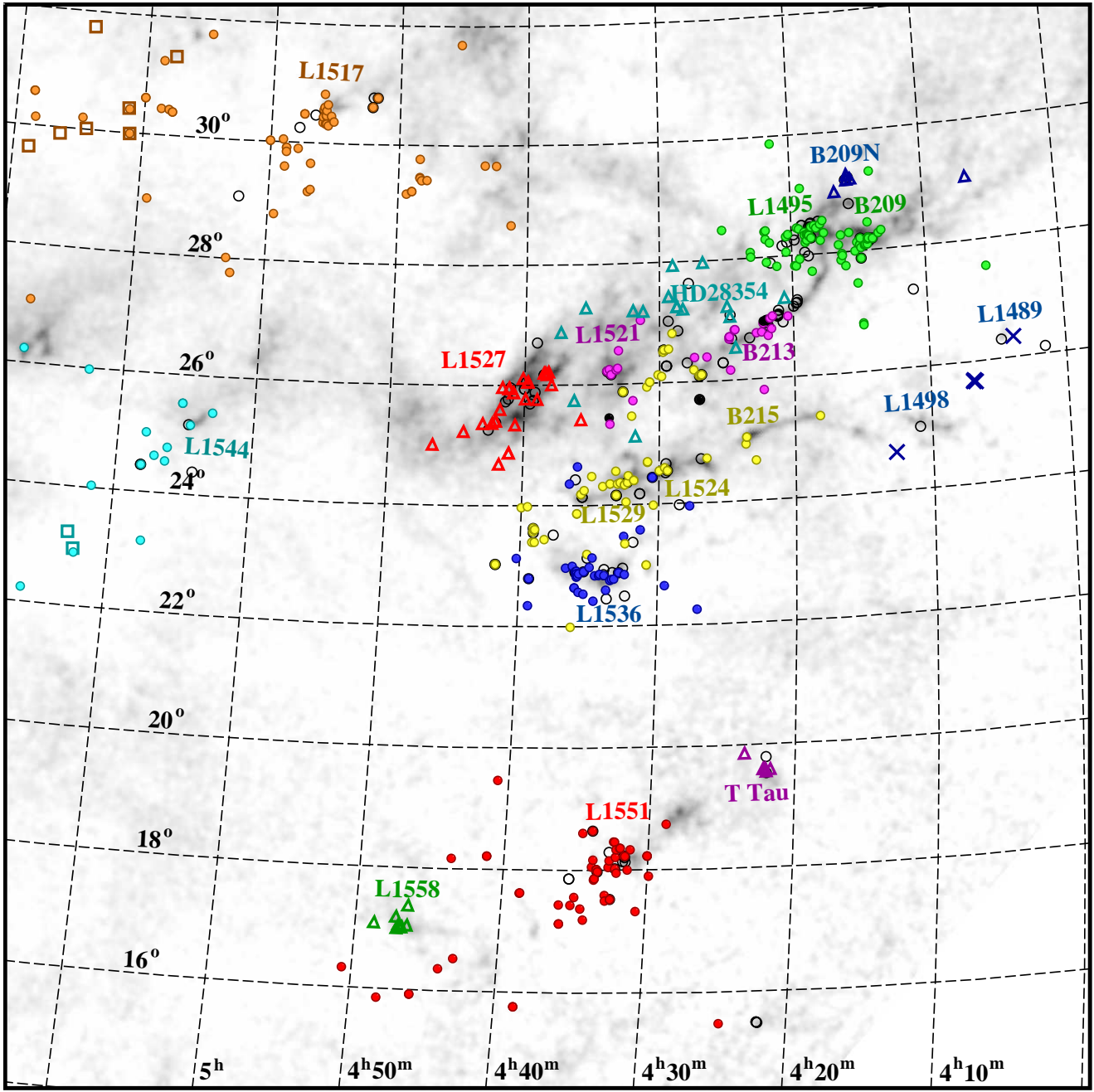}
\caption{
Equatorial coordinates for adopted members of Taurus (Table~\ref{tab:mem}).
The members with parallax measurements from Gaia DR3 are shown with filled
circles, open triangles, and crosses and the members that lack parallaxes
are plotted with open circles.
The former have been assigned to groups (and symbol types) based on their 
proper motion offsets (Figs.~\ref{fig:pp1}--\ref{fig:pp4}). 
A few Gaia-measured stars have uncertain group assignments (black filled 
circles). The map includes candidate members of the L1517 and L1544 groups
that lack spectra (open squares, Table~\ref{tab:cand}). The dark clouds in 
Taurus are displayed with a map of extinction \citep[gray scale;][]{dob05}.}
\label{fig:map1}
\end{figure}

\begin{figure}
\epsscale{1}
\plotone{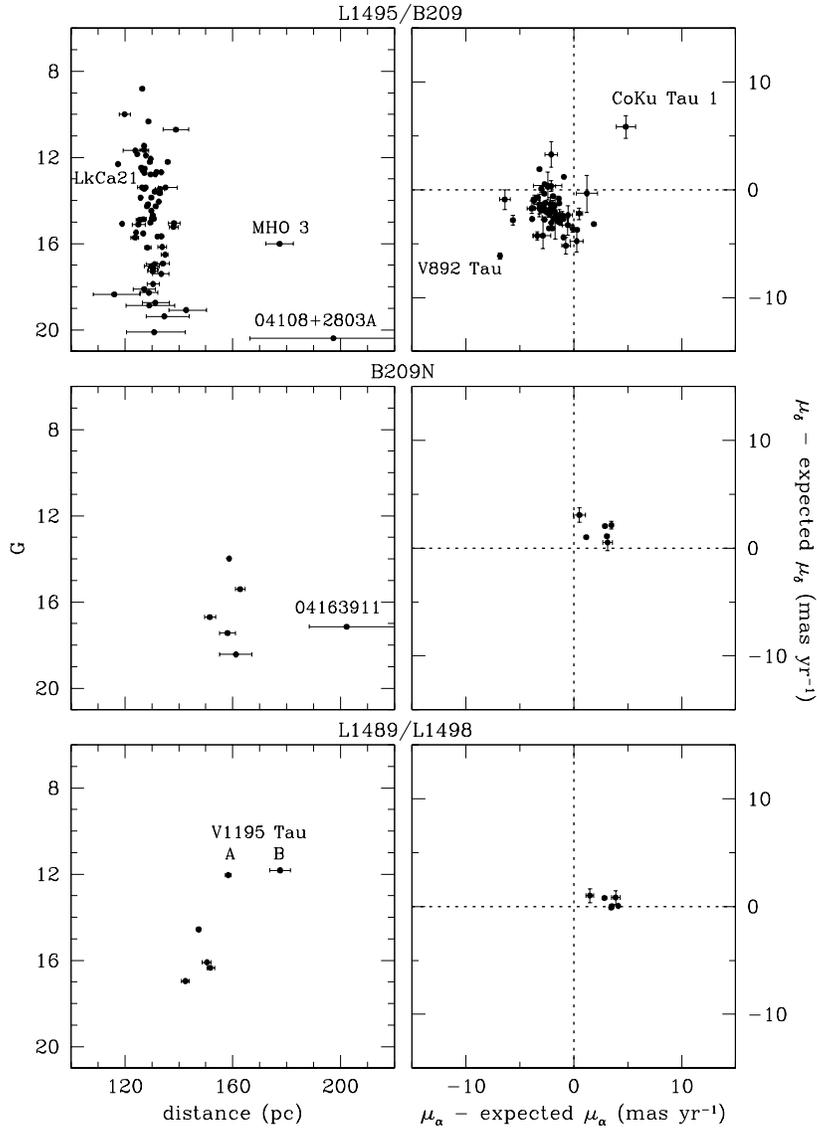}
\caption{
$G$ magnitudes, parallactic distances, and proper motion offsets based on
Gaia DR3 for members of L1495/B209, B209N, and L1489/L1498
in Taurus (Figure~\ref{fig:map1}).}
\label{fig:pp1}
\end{figure}

\begin{figure}
\epsscale{1}
\plotone{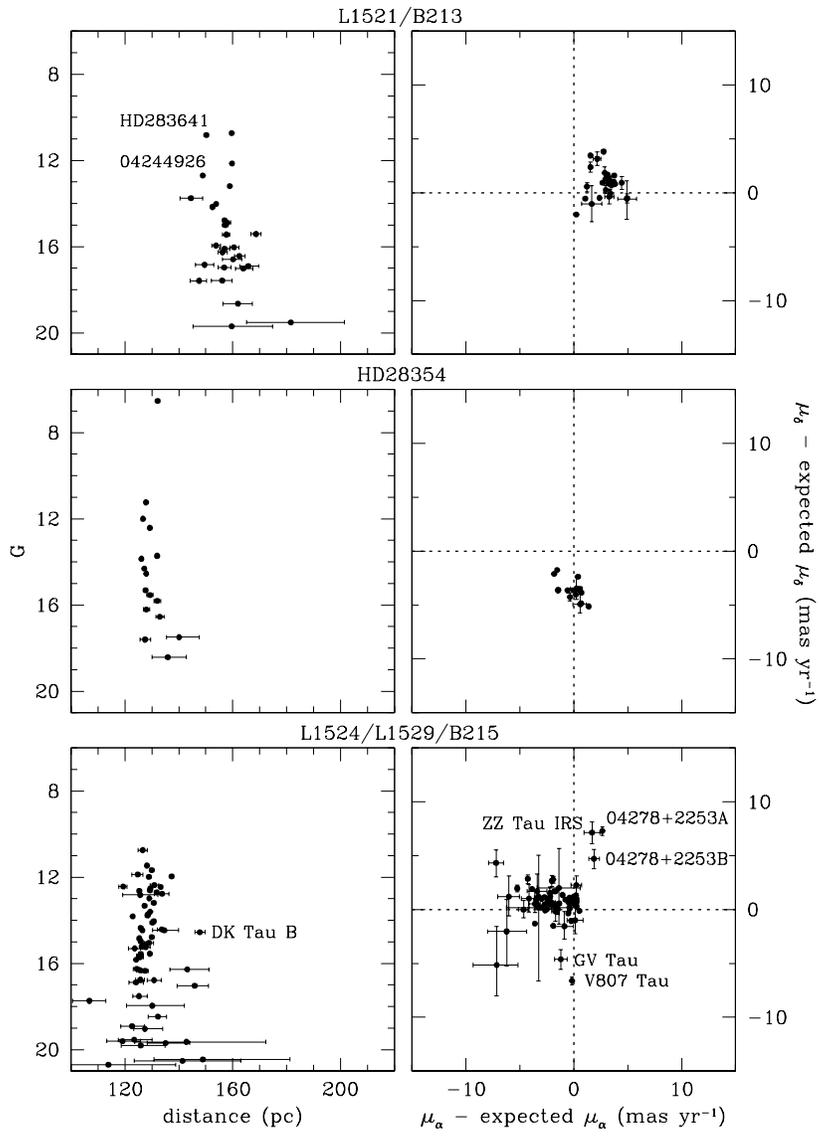}
\caption{
$G$ magnitudes, parallactic distances, and proper motion offsets based on
Gaia DR3 for members of L1521/B213, HD~28354, and L1524/L1529/B215
in Taurus (Figure~\ref{fig:map1}).}
\label{fig:pp2}
\end{figure}

\begin{figure}
\epsscale{1}
\plotone{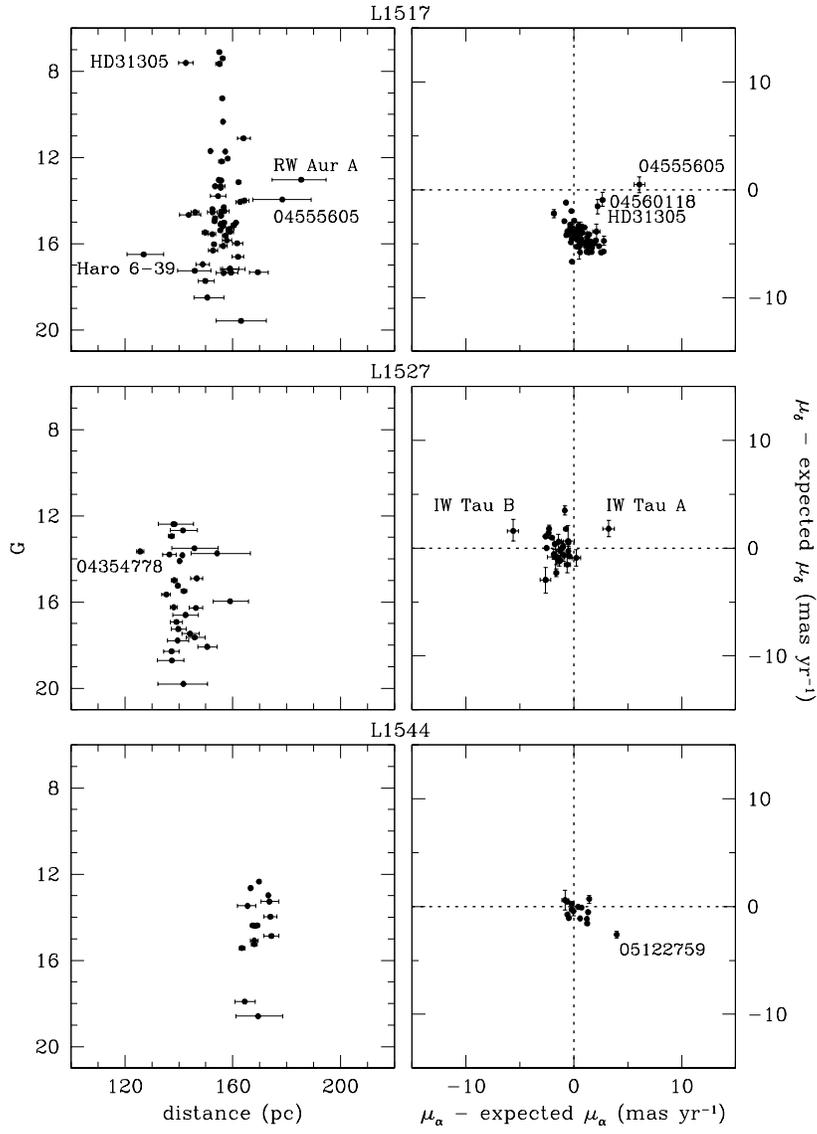}
\caption{
$G$ magnitudes, parallactic distances, and proper motion offsets based on
Gaia DR3 for members of L1517, L1527, and L1544 in Taurus 
(Figure~\ref{fig:map1}). 
}
\label{fig:pp3}
\end{figure}

\begin{figure}
\epsscale{1}
\plotone{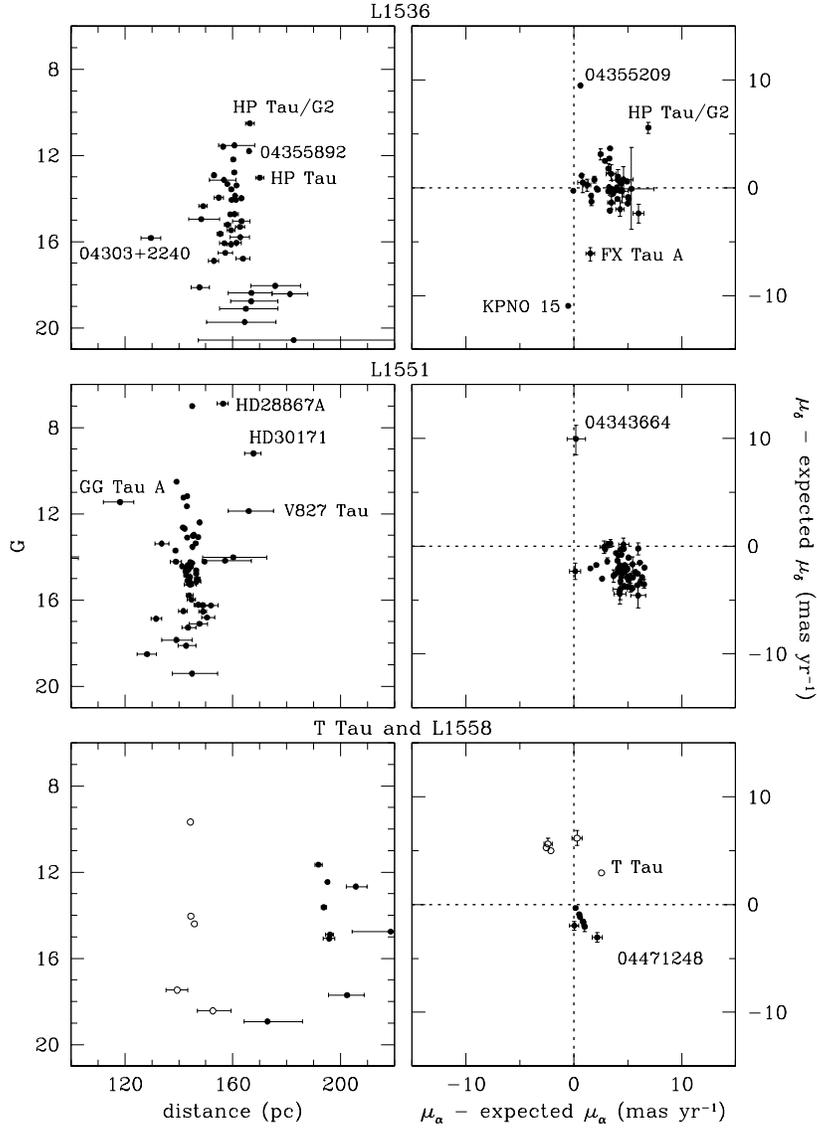}
\caption{
$G$ magnitudes, parallactic distances, and proper motion offsets based on
Gaia DR3 for members of L1536, L1551, L1558, and the T~Tau group in Taurus
(Figure~\ref{fig:map1}).}
\label{fig:pp4}
\end{figure}

\begin{figure}
\epsscale{1.1}
\plotone{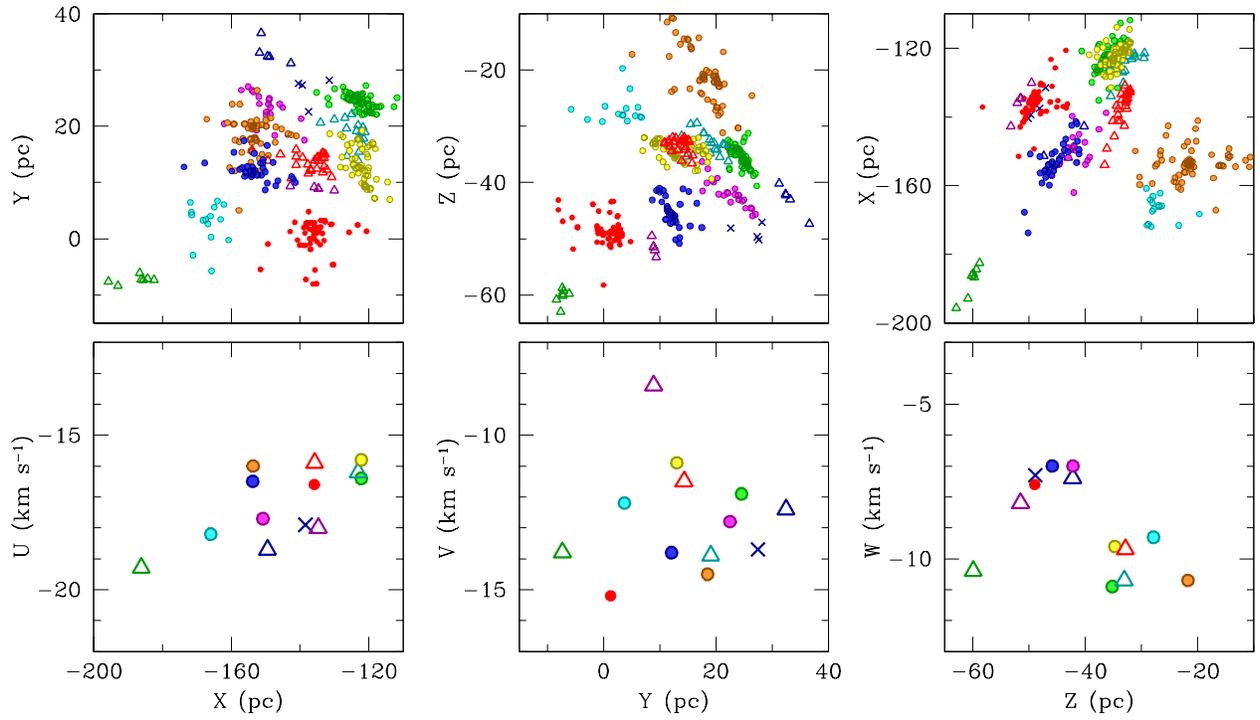}
\caption{
Galactic Cartesian coordinates for members of Taurus that have parallax
measurements (top). The median values of $UVW$ 
for the Taurus groups are plotted versus $XYZ$ (bottom).
The symbols are the same as those in Figure~\ref{fig:map1}.}
\label{fig:uvw}
\end{figure}

To illustrate the spatial clustering of the Taurus groups, 
I have plotted in the top row of Figure~\ref{fig:uvw}
the $XYZ$ positions in Galactic Cartesian coordinates for
the group members that have $\sigma_{\pi}<0.5$~mas using the symbols
in the map from Figure~\ref{fig:map1}.
The bottom row of Figure~\ref{fig:uvw} shows the three
components of the median velocities of the groups versus their
respective spatial positions. The groups exhibit a correlation between
$U$ and $X$ and an anti-correlation between $W$ and $Z$, which indicates
that portions of the complex are expanding in $X$ and contracting in $Z$. 

\begin{figure}
\epsscale{1.1}
\plotone{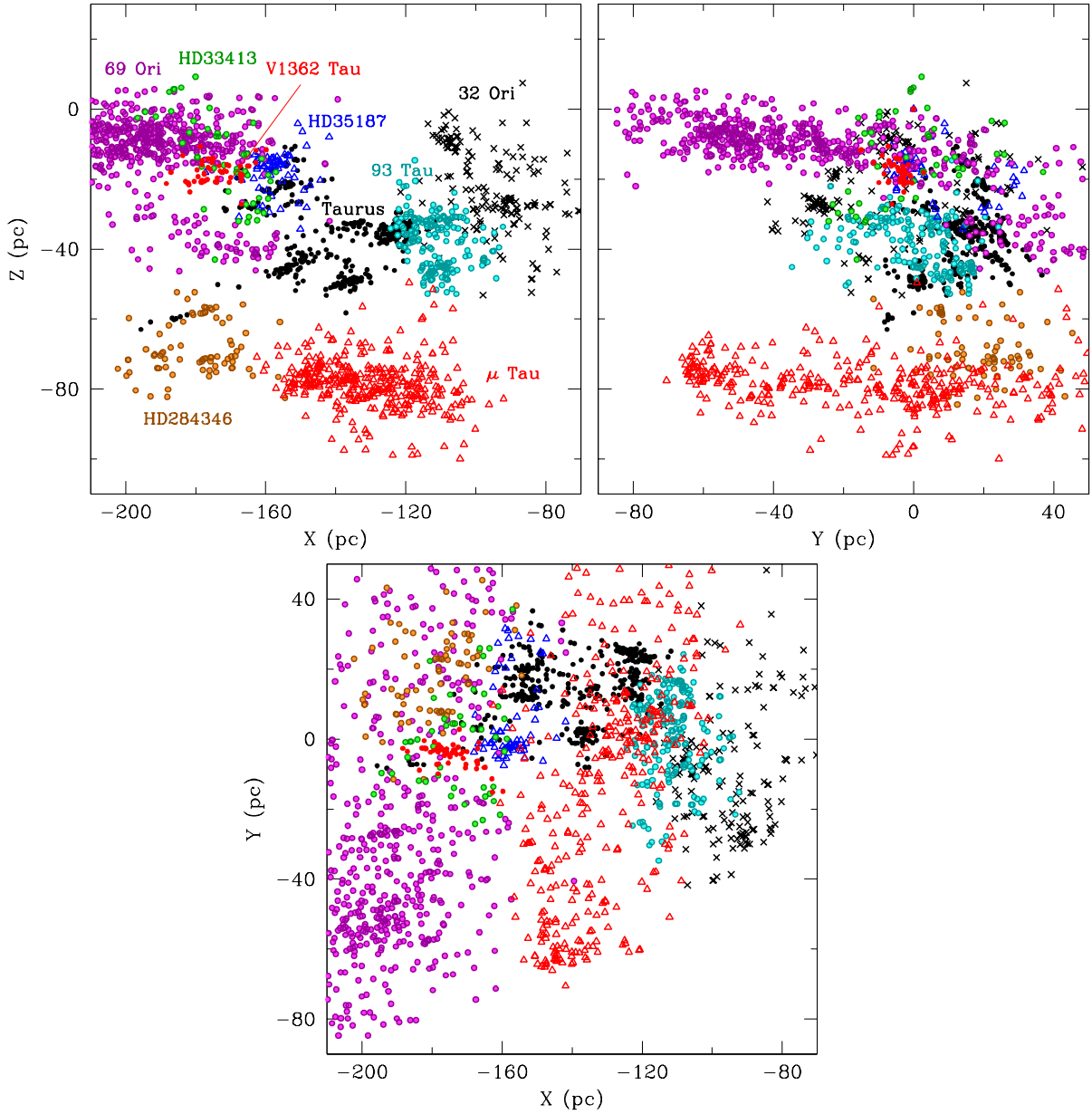}
\caption{Galactic Cartesian coordinates for members of Taurus 
(Table~\ref{tab:mem}) and candidate members of neighboring associations 
\citep[Table~\ref{tab:groups},][]{luh22o}.}
\label{fig:xyz}
\end{figure}

To compare the spatial distributions of the Taurus groups and the neighboring
associations, I have plotted in Figure~\ref{fig:xyz} the $XYZ$ positions
of their members using the symbols from the map in Figure~\ref{fig:map2}.
Although Taurus overlaps with several associations on the sky, most of the 
Taurus groups are fairly well isolated from the surrounding associations in 
spatial positions.  Since Taurus is near the Galactic anticenter, distance
is roughly along the $X$ axis. For instance, 32~Ori and 93~Tau are in the 
foreground of Taurus. The two largest associations, 69~Ori and $\mu$~Tau, are
broader in $X$ and $Y$ than in $Z$, i.e., they have sheet-like distributions 
that are parallel to the Galactic plane.

\begin{figure}
\epsscale{1.1}
\plotone{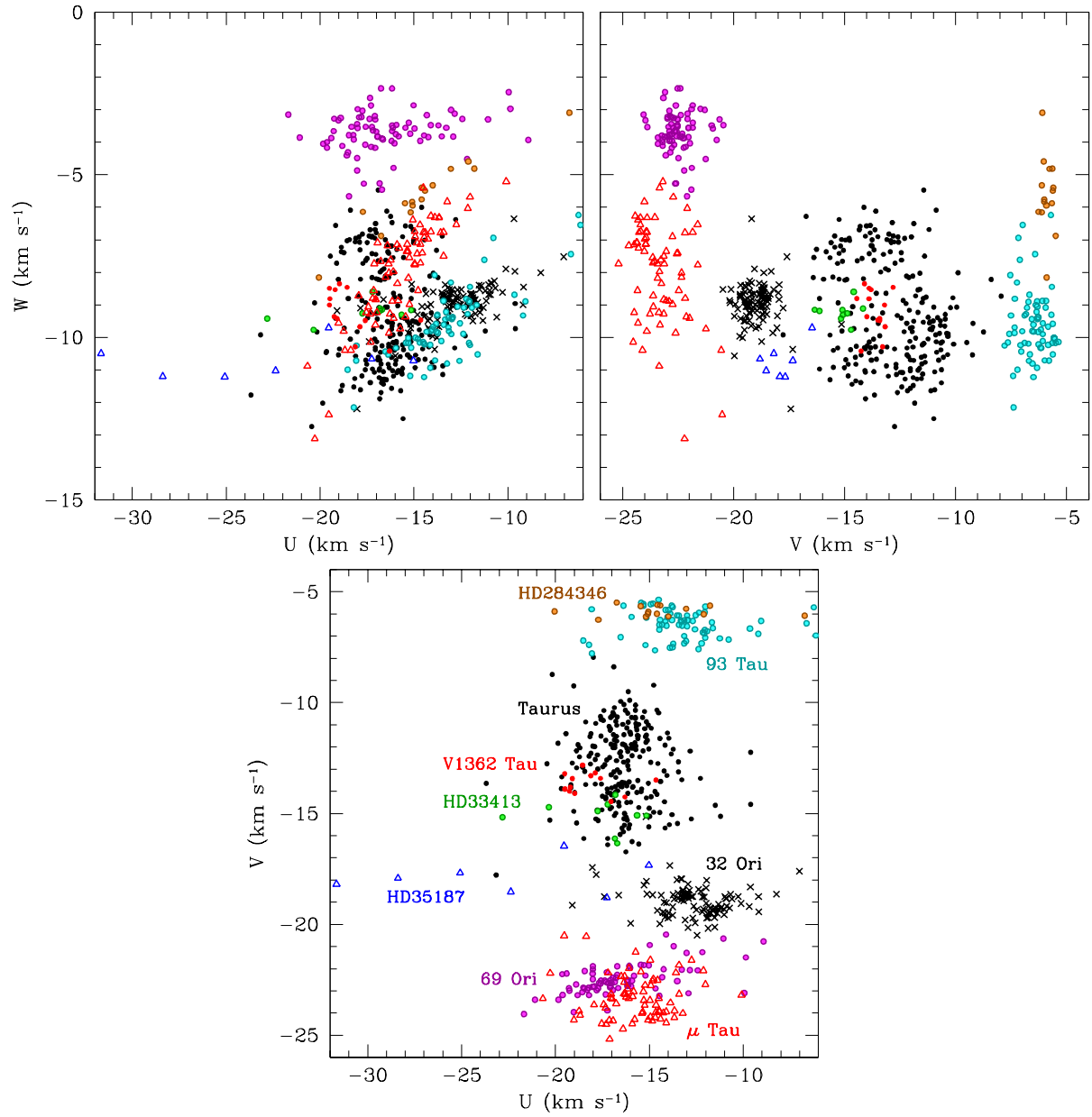}
\caption{$UVW$ velocities for members of Taurus (Table~\ref{tab:mem}) and 
candidate members of neighboring associations 
\citep[Table~\ref{tab:groups},][]{luh22o}.}
\label{fig:uvw2}
\end{figure}

Figure~\ref{fig:uvw2} shows the $UVW$ velocities that are available
in Taurus and the associations with the exception of 
the astrometric outliers and the stars with {\tt VSCATTER}$>3$~km~s$^{-1}$.
For most stars, the radial velocities contribute much larger errors
than the proper motions. Those large radial velocity errors
produce the stretching that is evident among the velocities in
Figure~\ref{fig:uvw2}. The angle of the stretching varies among the
associations because of their differing positions on the sky.
Taurus and the surrounding associations exhibited distinct kinematics
in terms of proper motion offsets in Figure~\ref{fig:pp5} and the same
is true for the $UVW$ velocities in Figure~\ref{fig:uvw2}.

\begin{figure}
\epsscale{1.1}
\plotone{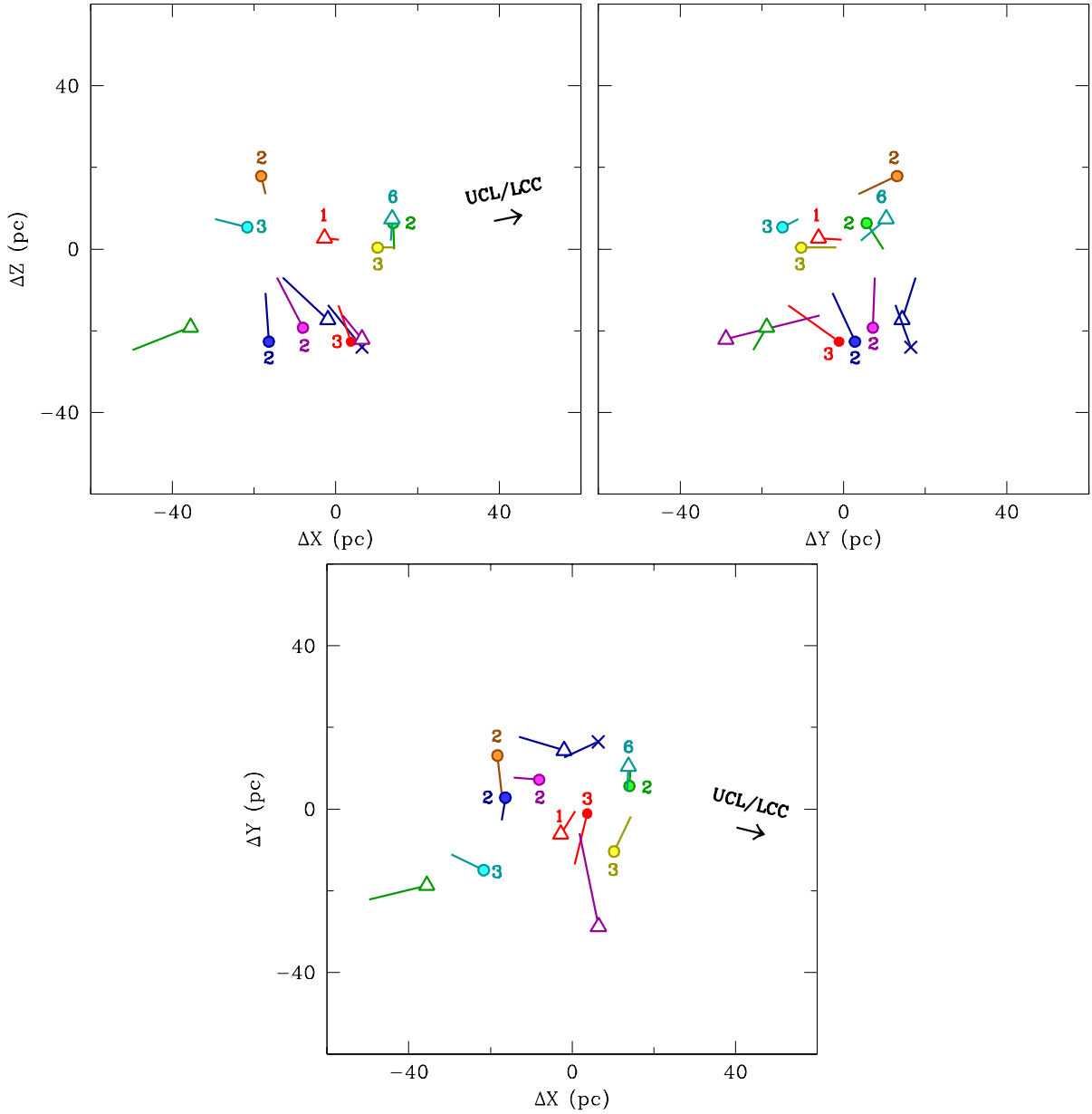}
\caption{
Offsets in Galactic Cartesian coordinates of the Taurus groups from the median 
position of Taurus members at 5~Myr in the past (symbols, 
Figure~\ref{fig:map1}).
The motion of each group during the last 5~Myr is represented by a solid line. 
The larger groups are labeled with their ages from Table~\ref{tab:ages}. 
The arrows indicate the direction of UCL/LCC 5~Myr ago.  
Since that direction is mostly along the $X$ axis, an arrow is not shown in
$\Delta Z$ versus $\Delta Y$.}
\label{fig:xyz2}
\end{figure}

Previous studies have proposed that the expansion of the Local Bubble was
propelled by a series of supernovae in UCL/LCC beginning 10--15~Myr ago 
\citep{mai01,fuc06,fuc09}, which resulted in the formation of many of the 
nearest molecular clouds, including Taurus, and the triggering of star 
formation within them \citep{zuc22}.  I have examined whether the positions 
and ages of the groups in Taurus show a pattern that would support that 
scenario. I have calculated the $XYZ$ positions of the Taurus groups
and UCL/LCC 5~Myr ago ($-$5~Myr) using their median positions and
velocities \citep[Table~\ref{tab:uvw},][]{luh22sc} and an epicyclic
approximation of Galactic orbital motion \citep{mak04}.
I selected $-$5~Myr because it is roughly similar to the oldest ages of
the Taurus groups. The results of this exercise are not sensitive to the
value that is adopted. In Figure~\ref{fig:xyz2}, I have plotted the offsets in
$XYZ$ of the Taurus groups from the median position of Taurus at $-5$~Myr
using the symbols from Figure~\ref{fig:map1}. I also include lines
that represent the motion of the groups over the past 5~Myr (i.e., the end
of a line that has no symbol is the current position of a group). 
The groups with age estimates from Table~\ref{tab:ages} are labeled with those 
ages after rounding to the nearest integer.  From the point of view of the 
Taurus clouds at that time, an expanding shell from UCL/LCC should have arrived 
from a direction between the direction of UCL/LCC at the time that the
supernovae began \citep[$-14$~Myr,][]{zuc22} and UCL/LCC's direction when the 
shell arrived at Taurus ($-5$~Myr). There is little difference between those 
two directions since most of the relative motion of UCL/LCC and Taurus is along 
the axis between them. The direction of UCL/LCC at $-5$~Myr is indicated in 
Figure~\ref{fig:xyz2}.

Assuming an expansion velocity of $\sim$8~km~s$^{-1}$ for the Local Bubble
at $-5$~Myr \citep{zuc22}, the expanding shell would have crossed the
Taurus clouds in $\sim$5~Myr. Thus, one might expect an age gradient 
of 5~Myr across the Taurus groups along the direction toward UCL/LCC.
Although the ages of the groups have large uncertainties, a gradient of that
size could be marginally detectable.
The oldest group, HD~28354 ($\sim$6~Myr), is indeed on the side of the 
Taurus complex facing UCL/LCC, but the two other groups on that side of
Taurus appear to have younger ages of 2--3~Myr. The remaining groups
that are farther from UCL/LCC have similar ages ($\sim1$--3~Myr).
It is possible that the interface between the interstellar medium and the 
Local Bubble was too clumpy and irregular for a clear correlation between
age and distance to appear.
Meanwhile, all of the associations in the vicinity of Taurus discussed in 
this work are too old for UCL/LCC to have played a role in their formation.

\citet{kra17}, \citet{ker21}, and \citet{kro21} have proposed that
previous samples of older stars near Taurus ($\gtrsim10$~Myr) have
a relationship with the Taurus clouds.
There are two possible forms for a relationship:
(1) the older stars were born within the existing Taurus clouds
or (2) they were born from clouds that were connected to the
gas that eventually formed the Taurus clouds, i.e., the Taurus groups
and the older stars represent different generations of star formation
within a cloud complex.
In the first scenario, the older stars should have the same average 
velocities as the younger stars associated with the Taurus clouds.
However, the stars from \citet{kra17}, \citet{ker21}, and \citet{kro21} that
were proposed as older members of Taurus do not share the same motions
with the Taurus groups based on their proper motion offsets and
$UVW$ velocities \citep[][Section~\ref{sec:compare}]{luh18}.
The same is true for the associations near Taurus,
as shown in Figures~\ref{fig:pp5} and \ref{fig:uvw2}.

In the second scenario, the neighboring clouds that produced the different
epochs of star formation should have motions that are similar or differ only
modestly, so the same would be true for the stellar populations that
they generated. Today, those populations would overlap spatially
or they would have drifted apart. In the latter case, the two populations should
still trace back to adjacent locations in the past.
For each association near Taurus, I have calculated its distance from each 
of the Taurus groups (or its progenitor gas) when the association was born 
using the median positions and velocities of the association and the Taurus 
groups and the epicyclic approximation of Galactic orbital motion.
The V1362~Tau association (13~Myr) traces back to within $\sim7$~pc from the
past location of the L1527 group, which is near the center of Taurus.
None of the other associations had birth sites within 30~pc of any of the
Taurus groups. Although the HD~35187 association overlaps with one of the 
Taurus groups on the sky (L1517, Section~\ref{sec:ker21}), their differing 
velocities indicate they would have been widely separated ($>$100~pc) when 
HD~35187 was born. Thus, V1362~Tau is the only one of the associations that
could have a relationship with Taurus. 
It is not possible to conclusively assess that possibility.
Given the number of associations in the vicinity of Taurus, 
a close approach between a cloud complex with the size of Taurus
and an unrelated association is plausible.

When assessing whether the Taurus groups belong to a complex that
has produced earlier episodes of star formation, it is useful to
consider the properties of known complexes of that kind.
Two of the nearest examples of molecular clouds that are related to older
generations of stars are Corona Australis and Ophiuchus.
In both cases, (1) the cloud and its associated stars are located within
a richer and more extended distribution of older stars (10--15~Myr,
Upper Corona Australis and Upper Sco)
and (2) the older population spans a broader range of velocities that
encompasses the narrower range of velocities of the younger stars
\citep{luh20u,luh22sc,esp22}.
An older population with those characteristics is not present in the 
Taurus complex.

\section{Comparison to Recent Studies}
\label{sec:compare}

\subsection{\citet{gal19}}

\citet{gal19} applied a hierarchical clustering algorithm
to the equatorial coordinates, proper motions, and parallaxes of 519
candidate Taurus members from \citet{jon17} and \citet{luh18}
that had measurements of proper motions and parallaxes from Gaia DR2
or radio interferometry. My catalog of adopted members includes 485
of those stars. I classify the remaining 34 stars from \citet{gal19} as
likely nonmembers. The analysis from that study produced 21 clusters,
half of which contain only 2--5 stars per cluster.
Two of those small clusters,\#2 and \#5, are entirely nonmembers.
For seven clusters, their cluster corresponds to a subset of a single group
from Section~\ref{sec:assign} (i.e., the two studies identify the same
groups). However, five of my groups (L1517, L1521/B213, L1524/L1529/B215,
L1527, HD~28354) are each broken into 2--4 clusters in \citet{gal19}, 
which is likely partially due to the fact that their clustering analysis did 
not account for the projection effects that can cause stars with similar 
velocities to have different proper motions, particularly in a population 
like Taurus that is widely distributed on the sky. Even with corrections
for projected effects, a clustering analysis performed in part on celestial
coordinates is susceptible to breaking an association into small fragments 
if it is clumpy \citep[][]{gag20}.
Finally, one of their clusters (\#7) is a mixture of two of my groups,
L1495/B209 and HD~28354. These two groups are adjacent spatially, but
have different kinematics and ages (Section~\ref{sec:pops}).

\subsection{\citet{roc20}}

\citet{roc20} used a maximum likelihood technique to fit the
Gaia DR2 proper motions and parallaxes of a selection of 283 candidate members
of Taurus from \citet{esp19} in terms of multiple populations, each of which
was defined by means and dispersions in parallax and proper motion
and a normalization factor that accounted for the fraction of all stars 
in the population. Their analysis produced six populations.
Five of their populations contain stars that span three or more of the Taurus
groups defined in this work. For some of the populations, most notably
``Taurus D", the stars are projected against multiple clouds across a
large area of Taurus and exhibit significant variations in kinematics
as measured by their proper motion offsets and space velocities.
These discrepancies likely arise from the fact that the fitting of the
populations was performed on proper motions without correction for
projection effects. For instance, stars in different Taurus clouds
can have similar proper motions but different space velocities because
of those effects. As a result, most of the populations from \citet{roc20}
do not represent coherent stellar groups.

\subsection{\citet{liu20,liu21}}

\citet{liu20} found two new associations, u~Tau and e~Tau,
southwest of the Taurus clouds based on the clustering of their members
in equatorial coordinates, proper motions, and parallaxes from Gaia DR2.
The first association is sufficiently far from Taurus that I have not
attempted to identify its members. The second one has been named 
$\mu$~Tau by \citet{gag20}. My sample for $\mu$~Tau includes 84 of the
119 candidates from \citet{liu20}.

\citet{liu21} used data from Gaia DR2 to search for groups of young stars
($<$100~Myr) within a large volume extending well beyond the Taurus clouds.
The groups were identified via clustering in spatial positions and tangential
velocities that were calculated from proper motions and parallactic distances.
Their analysis produced 22 groups.
Groups 1--8 from \citet{liu21} contain 277 stars; 246 are among my
adopted members of Taurus and 31 are classified in this work
as field stars or candidate members of associations near Taurus.
The remaining groups from that study correspond to subsets of the
associations that were described in Section~\ref{sec:groupsearch}.
The group number from \citet{liu21}, the number of stars from that group
that are within an association from Section~\ref{sec:groupsearch}, and the
name of the latter are as follows:
group 9, 40 stars, and V1362~Tau; group 10, 29 stars, and HD~35187;
group 11, 32 stars, and 32~Ori; groups 12--14, 101 stars, and 93~Tau;
groups 15--21, 96 stars, and 69~Ori; and group 22, 15 stars, and $\mu$~Tau.
Groups 12--14 also included 77 of the candidate members of 93~Tau (group 29)
from \citet{luh18}.
The multiple groups that \citet{liu21} identified for 93~Tau and 69~Ori
are adjacent to each other on the sky, which is likely a reflection of the
fact that the analysis from that study detected clusters in tangential 
velocities, which vary with celestial coordinates for a given space velocity.

\subsection{\citet{gag20}}

\citet{gag20} identified candidate members of the $\mu$~Tau association 
using astrometry and photometry from Gaia DR2 in conjunction with a model
for the spatial positions and velocities of members of the association.
They arrived at 393 higher-quality candidates and 155 lower-quality candidates,
208 and eight of which appear in my sample of candidates for $\mu$~Tau,
respectively. Most of their candidates that are absent from my sample were
rejected in my analysis via kinematics and CMDs.  A few do satisfy those
criteria, but are outside the range of distances in which I selected candidates.
\citet{gag20} included white dwarf candidates in their sample, which I did
not attempt to identify (Section~\ref{sec:ident}). My catalog contains 138
candidates that are not in the sample from \citet{gag20}.

\subsection{\citet{kra17}}

\citet{kra17} compiled 396 diskless stars that had been previously identified
as possible members of Taurus. They classified 218 of those stars as
confirmed or likely members of Taurus based on diagnostics of youth
and pre-Gaia kinematic data. 
Roughly 40\% of the proposed members were absent from my prior catalogs
of Taurus \citep[e.g.,][]{luh17}, most of which were older and more widely
scattered than the adopted members from the latter study.
\citet{kra17} suggested that those stars represented an older distributed
population that was associated with the Taurus complex.
The high-precision parallaxes and proper motions from Gaia DR1 and DR2
made it possible for \citet{esp17} and \citet{luh18} to closely scrutinize
the membership of the older candidates from \citet{kra17}.
For the candidates that were located within the field considered by \citet{luh18}
and that had kinematic measurements from Gaia, none were comoving with the 
groups of stars associated with the Taurus clouds. A subset of those stars 
shared similar motions and spatial positions with group 29 from 
\citet{oh17} (93~Tau in this work), which were included in a new catalog
for that association by \citet{luh18}.

I have performed an updated examination of the candidates from \citet{kra17}
with my new survey of Taurus and its neighboring associations with Gaia DR3.
I find that their 218 proposed members of Taurus included a mixture of
stars from the Taurus groups (126), 93~Tau (26), V1362~Tau (8),
and 32~Ori (11). Two candidates lack parallax measurements from Gaia.
The remaining 45 stars have kinematics that are inconsistent with
Taurus and that span a wide range. A few of those stars are roughly
comoving with the associations near Taurus, but are located sufficiently far
from the latter that membership is uncertain. For instance, 15 of those 45
candidates have distances of $<$100~pc or $>$200~pc.
As in \citet{luh18}, I find that the older candidates from
\citet{kra17} are not comoving with the Taurus groups, and instead are
isolated field stars and members of the associations near Taurus.

\subsection{\citet{ker21}}
\label{sec:ker21}

\citet{ker21} used data from Gaia DR2 to search for groups of
young low-mass stars in the solar neighborhood via CMDs and clustering
in spatial positions and tangential velocities.
Eleven of the resulting groups were near the constellation of Taurus, so 
they were named with the affix of ``GT" for ``Greater Taurus".
Each group included a core sample and an expanded sample
that was expected to have greater contamination from nonmembers.
Only the core samples are considered in the following discussion.

\begin{figure}
\epsscale{1.1}
\plotone{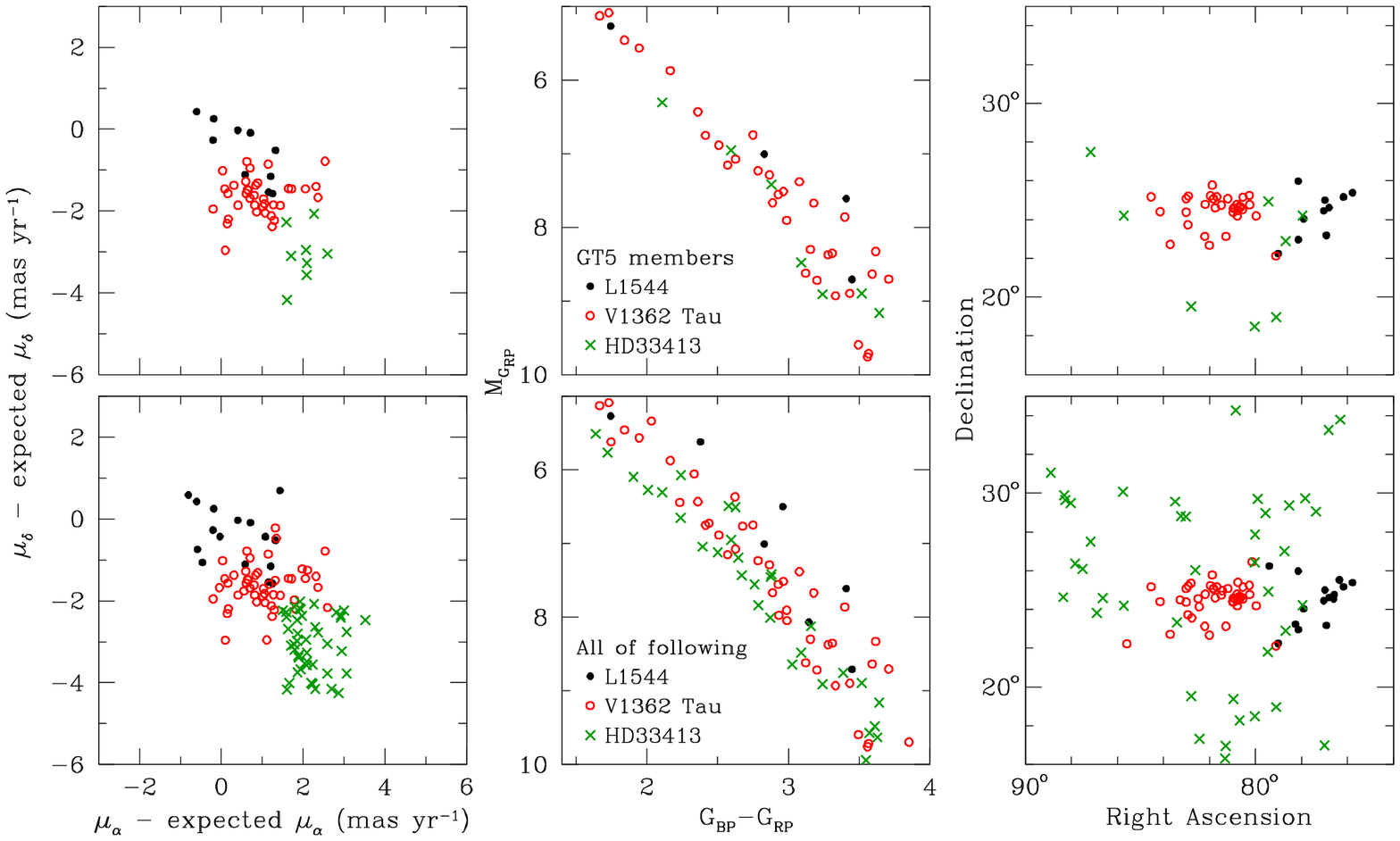}
\caption{
Top: Proper motion offsets, CMDs,
and equatorial coordinates for members of the GT5 group from \citet{ker21},
which are assigned to three different groups in this work, as indicated by
the symbols. Bottom: All candidate members of those three groups from this work.
Stars with full disks have been omitted from the CMDs.
}
\label{fig:tri}
\end{figure}

\begin{figure}
\epsscale{1.1}
\plotone{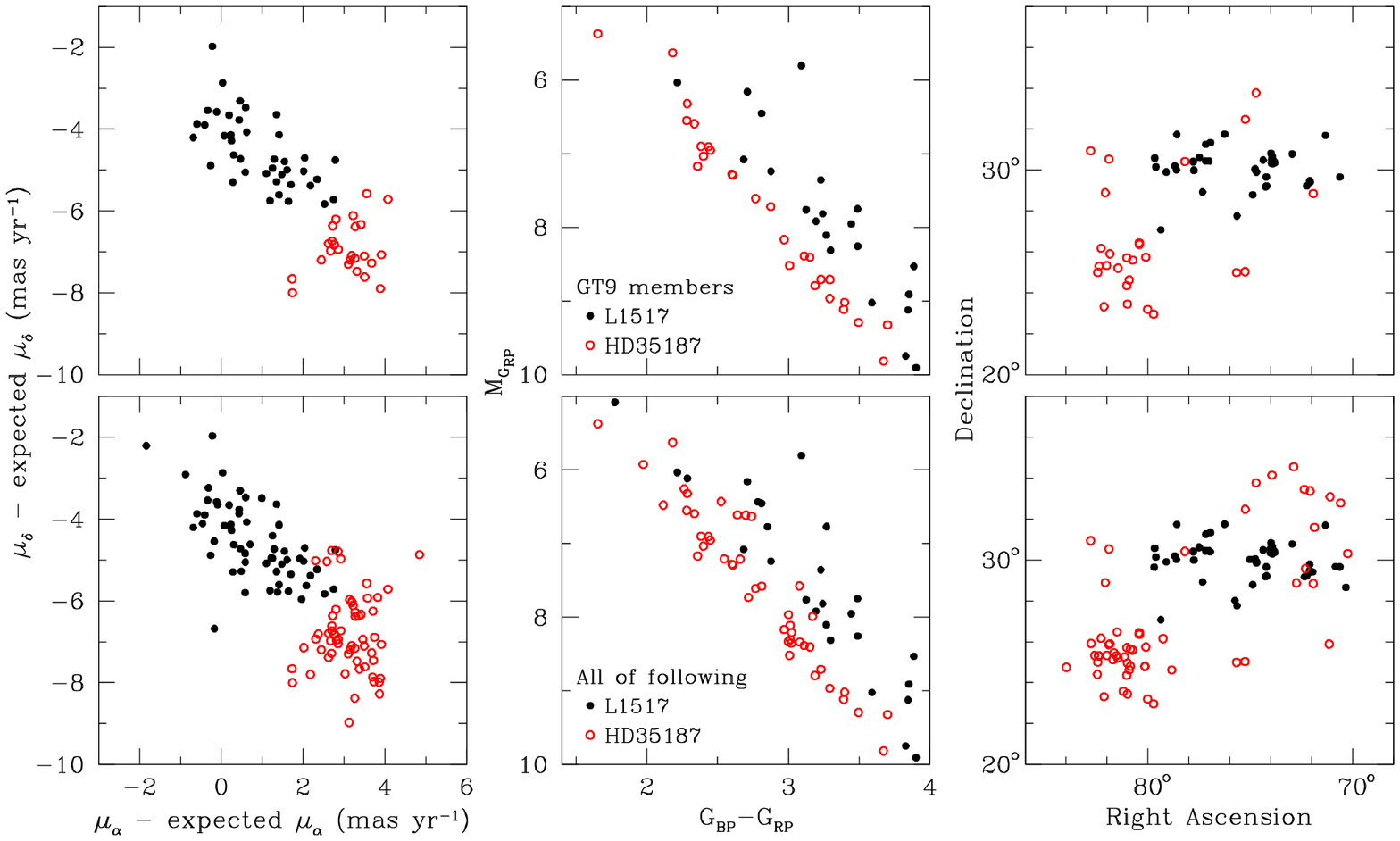}
\caption{
Top: Proper motion offsets, CMDs,
and equatorial coordinates for members of the GT9 group from \citet{ker21},
which are assigned to two different groups in this work, as indicated by
the symbols. Bottom: All candidate members of those two groups from this work.
Stars with full disks have been omitted from the CMDs.
}
\label{fig:tri2}
\end{figure}

As recognized by \citet{ker21}, GT1 (17 stars) is a small fragment of the
$\mu$~Tau association. 
GT2 (25 stars) is quite far from the nearest Taurus cloud ($\sim25\arcdeg$), so 
I have not attempted to identify its members.  
GT3 (46 stars) consists of three distinct clusters in spatial positions and 
proper motion offsets that correspond to $\mu$~Tau and two much younger
groups that are more than $10\arcdeg$ south of the Taurus complex.
The latter two groups were identified as GT3A and GT3B by \citet{ker21}
and are sufficiently far from Taurus that I have not performed a census of them.
GT4 (51 stars) is dominated by members of 93~Tau, including 21 candidates
from \citet{luh18} and 39 candidates from this work.
According to my membership classifications, 
GT5 (55 stars) and GT9 (77 stars) each consist of a combination of a Taurus
group and one or more older associations, all of which have distinct 
kinematics, spatial positions, and ages and hence do not belong in the same
populations. GT5 is a mixture of L1544 in Taurus and two older associations, 
V1362~Tau and HD~33413. GT9 combines members of L1517 in Taurus with the older
association HD~35187. To illustrate the differing kinematics, positions, and
ages of the groups within GT5 and GT9, I present in Figures~\ref{fig:tri}
and \ref{fig:tri2} the proper motion offsets, CMDs, and equatorial
coordinates for (1) the members of GT5 and GT9, which are plotted with
symbols that indicate the groups to which they are assigned in my analysis
and (2) all candidates that I have identified for those groups.
Only stars that lack full disks are shown in the CMDs so that the comparisons
of ages are not affected by disk-related phenomena (Figure~\ref{fig:cmd1}). 
As discussed in \citet{luh22o}, GT6 and GT7 (33 and 11 stars) are fragments
of 32~Ori. The remaining three groups from \citet{ker21}, GT8, GT10, and GT11,
correspond to groups in Taurus. GT8 (87 stars) combines members of four groups 
(L1495/B209, L1524/L1529/B215, L1527, HD~28354), GT10 (30 stars) is a subset
of L1551, and GT11 (34 stars) includes members of three groups (B209N, 
L1521/B213, L1536). As with some of the previously mentioned studies, the 
identification of groups via clustering in tangential velocities in 
\citet{ker21} was prone to both breaking extended associations into small 
fragments (as noted by that study) and assigning stars with different
space velocities to the same group.

\subsection{\citet{kro21}}

To analyze the properties of groups within Taurus, \citet{kro21}
compiled 571 Gaia-detected stars that they considered to be candidate
members of Taurus based on previous surveys.
My catalog of adopted members includes 452 of their candidates.
Among the 119 stars from \citet{kro21} that are absent from my catalog,
53 are classified as members of the older associations near Taurus in 
Section~\ref{sec:groupsearch}, 59 were rejected as candidates for
Taurus or the older associations in Section~\ref{sec:groupsearch},
and seven lack parallax measurements. 
The 59 rejected stars include 10 that have distances of
$<$100~pc or $>$200~pc, and thus are far from the Taurus clouds.
The seven stars that lack parallaxes consist of
HD~284135, RX~J0420.8+3009, Gaia DR3 3411198987968134528, 152416436441091584,
3415706130945884416, 149367387618890624, and 149370651795329536.
The first three are far from the Taurus groups on the sky, so parallax
data are needed to assess whether they are members of Taurus or the neighboring
associations.
Gaia DR3 152416436441091584 and 3415706130945884416 are possible
companions to adopted Taurus members \citep{esp19}, but I have not
included them in my catalog since they lack any data that could better
constrain companionship, such as parallaxes, colors, and spectra.
Gaia DR3 149367387618890624 and 149370651795329536 were matched to GV~Tau~B
and IRAS~04264+2433 by \citet{kro21}, respectively, but the position of the
first Gaia source relative to GV~Tau is inconsistent with
previous astrometry of the pair \citep{lei93} while the second Gaia source
appears to be nebulosity (Section~\ref{sec:adopt}).

Many of the stars in the catalog from \citet{kro21} that are absent from
my census of Taurus have ages of $\gtrsim$10~Myr in CMDs.
\citet{kro21} proposed that those older stars have a relationship with
the Taurus complex. In Section~\ref{sec:ages}, I examined that possibility
for the associations near Taurus, which contain 53 of the stars from
\citet{kro21}, as mentioned earlier in this section.  I now perform a similar 
analysis for the 59 stars that I rejected for membership in Taurus or the 
older associations.
The proper motion offsets of those stars span a large range, indicating
that the stars do not comprise a coherent stellar population (e.g., an earlier 
generation of stars associated with Taurus) and instead are likely to be young 
field stars. Among the 49 stars at distances of 100--200~pc, 36 have 
measurements of radial velocities, enabling estimates of $UVW$ velocities.
Using those velocities, the ages implied by CMDs, and the epicyclic
approximation of Galactic orbital motion, I have estimated the $XYZ$
positions of those 36 stars when they were born. Only two of the
stars with ages of $\gtrsim10$~Myr were located within 10~pc of
the positions of the Taurus groups (or their progenitor gas) at those times.
Thus, that sample contains no evidence of an older population of stars
associated with the Taurus complex. One of the older stars in question is the 
spectroscopic binary St34. It was included in catalogs of Taurus members at 
one time \citep[e.g.,][]{ken95}, but its membership was questioned 
by \citet{har05} based on the relatively old age implied by the absence of
Li absorption \citep{whi05}.
Its $UVW$ velocity differs by $>4$~km~s$^{-1}$ from median velocities of all
of the Taurus groups and it differs by $\sim9$~km~s$^{-1}$ from that of the
nearest group, L1551, indicating that St34 is not a member of any of the Taurus 
groups. Based on my traceback calculations,
St34 was separated by $\gtrsim80$~pc from the progenitor clouds of the
Taurus groups when it was born, which demonstrates that St34 has no 
relationship with Taurus. Instead, St34 and the other stars discussed here are
examples of the numerous young stars that permeate the solar neighborhood
and whose parent associations have either dissolved or are too diffuse
to be recognizable.

\citet{kra17}, \citet{ker21}, and \citet{kro21} described
their proposed older members of Taurus as a ``distributed" population,
which typically refers to widely scattered stars that surround and encompass
a clustered population.
However, many of those stars are members of the associations near Taurus,
such as 32~Ori and 93~Tau, and my work has demonstrated that those
associations are spatially separate from the Taurus groups
(Figure~\ref{fig:xyz}) and thus do not represent a distributed population
relative to Taurus.
Meanwhile, as discussed previously, the remaining older stars span a wide
range of kinematics and are very likely to be young field stars, which
naturally have a distribution that is uniform and widely scattered.

\citet{kro21} sought to identify groups within Taurus by applying a Gaussian
mixture model (GMM) to the Galactic Cartesian coordinates of their adopted
members that had parallax measurements from Gaia DR3. They adopted a model
with 14 components, which were labeled as C1--C10 and D1--D4.
My group classifications for the members of those components are summarized
in Table~\ref{tab:kro}. A few stars are listed as ``no group" because
their group assignments are uncertain (Section~\ref{sec:assign}).
For some of the components from \citet{kro21}, there is a direct 
correspondence to a single one of the groups that I have defined, which 
occurs for groups that are spatially isolated.
However, several of the components contain members of multiple groups 
from my work that have varying kinematics, which results from the fact that
kinematics were not considered by \citet{kro21} in their GMM model.
As an example, C1 consists of the L1551 and T~Tau groups, which do not share 
the same kinematics and do not comprise a single stellar population, as shown 
in Figure~\ref{fig:pp4}. In addition, some of the components (e.g., D2, D4) 
combine several Taurus groups, older associations near Taurus, and stars that 
are not members of either Taurus or known associations (young field stars).
As a result, those components are not coherent stellar populations
and hence do not have physically meaningful properties.

\subsection{Membership of Planet-hosting Young Stars}

\citet{dav19} detected a Jupiter-sized planetary companion to V1298~Tau.
It was among the nine candidate members of group 29 from \citet{oh17} (the 93 
Tau association), which were identified with the first data release of Gaia.
It was not included in the larger sample of 91 candidates selected with
DR2 by \citet{luh18} because it was a modest outlier in parallax.
V1298~Tau does appear in my new sample of 192 candidates for 93~Tau, so the age 
for that association can be adopted for V1298~Tau
(35~Myr, Section~\ref{sec:ages}).

\citet{yu17} reported evidence of a close-in gas giant around V1069~Tau 
(TAP~26, HBC~376). Because it is located within a few degrees of T~Tau, 
\citet{yu17} adopted the distance of T~Tau for V1069~Tau, arriving at an age 
estimate of 17~Myr based on its position in the Hertzsprung-Russell and
the predictions of evolutionary models.  However, the Gaia DR3 parallax 
for V1069~Tau corresponds to a distance of 121~pc, resulting in an older age.
In addition, I find that the star is kinematically distinct from Taurus and
is a candidate member of 93~Tau.

In direct imaging, \citet{gai22} resolved a planetary-mass companion
at a separation of $0\farcs9$ from 2MASS J04372171+2651014 (hereafter 2M0437), 
which was identified as a Taurus member by \citet{esp17} and is classified in 
this work as a member of the Taurus group associated with HD~28354 
(6~Myr, Section~\ref{sec:ages}).
To search for additional components of this system, \citet{gai22}
checked Gaia DR3 for sources within an angular distance of $1000\arcsec$
that are comoving with the primary. They found one possible companion,
2MASS J04372631+2651438 (hereafter 2M043726), which has
a separation of $75\arcsec$, a similar proper motion to 2M0437 
(differing by $\sim2$~mas~yr$^{-1}$), and an uncertain parallax
($8.3\pm2.8$~mas) that is consistent with that of 2M0437 ($7.81\pm0.03$~mas).
Based on those data, \citet{gai22} concluded that 2M043726 is a Taurus member
and is likely to be a companion to 2M0437 (as opposed to an unrelated member).
They measured a spectral type of K5--M3 for 2M043726 using near-IR
spectroscopy, which is earlier than expected for a Taurus member given its 
faint photometry (i.e., it is fainter than expected for a Taurus member near
that spectral type). This is reflected in the fact that 2M043726 appears
well below the sequence of Taurus members when placed in a diagram of
$M_{G_{\rm RP}}$ versus $G-G_{\rm RP}$ using the distance of 2M0437.
To explain the anomalously faint photometry of 2M043726, \citet{gai22} 
proposed that it is occulted by an edge-on disk. They concluded that the 
available mid-IR images are inadequate to verify that a disk is present.

I have assessed the companionship of 2M043726 with 2M0437 and
its membership in Taurus.
Although the 1~$\sigma$ errors for the parallax of 2M043726 overlap with
the parallax of 2M0437, the same is not true for the distances inferred
from those parallaxes by \citet{bai21}, which are 
$270^{+137}_{-106}$~pc for 2M043726 and $127.2^{+0.3}_{-0.4}$~pc for 2M0437.
Thus, 2M043726 is unlikely to be a companion to 2M0437 or a member of
its Taurus group based on those distance estimates.
For Gaia DR3 sources within $3\arcdeg$ from 2M0437 that are not known
Taurus members, the average number of stars with proper motions similar to 
that of 2M0437 ($\Delta\mu\lesssim3$~mas~yr$^{-1}$) and unconstrained
values of parallax corresponds to $\sim1.7$ for an area equal to the
field searched by \citet{gai22}, so it is not surprising that their
search would find a field star that appeared to be comoving with 2M0437.
Meanwhile, the proposal that 2M043726 has an edge-on disk can be tested
with mid-IR photometry from the Spitzer Space Telescope.
2M043726 was detected by Spitzer in four bands from 3.6--8.0~\micron\ and was
not detected in a band at 24~\micron. The 3.6--8.0~\micron\ detections
are consistent with photospheric emission and do not show excess emission
from a disk. In comparison, 
several known members of Taurus have edge-on disks, all of which have excess
emission at 8.0~\micron\ (and at shorter wavelengths in some cases) and
have large enough excesses at 24~\micron\ that they are easily detected
in that band \citep{luh10}. Therefore, 2M043726 is unlikely to have an edge-on 
disk, in which case its photometry is inconsistent with Taurus membership.
Finally, I note that 2M043726 lacks any evidence of youth, which is desirable 
when assigning membership in Taurus. The available data indicate that 
2M043726 is not a member of Taurus and therefore is not a companion to 2M0437.

\section{Conclusions}

I have used high-precision photometry and astrometry from Gaia DR3 
to perform a census of the Taurus star-forming region and young associations 
in its vicinity. The results are summarized as follows:

\begin{enumerate}

\item
I have used proper motions and parallaxes from Gaia DR3 to vet my
previous census of Taurus for nonmembers \citep{esp19} and to search 
for new members, which has resulted in minor updates.
The new catalog contains 532 adopted members.
I have compiled various data for these sources, including spectral
classifications, Gaia photometry and astrometry, and radial velocities.
Measurements of parallaxes with $\sigma_{\pi}<1$~mas and radial velocities
are available for 412 and 330 members, respectively.
In addition to the adopted members, there remain nine candidates
identified with Gaia DR3 that lack spectral classifications.

\item
The Taurus complex contains multiple clouds and associated stellar groups,
which have modestly different average velocities.
For adopted Taurus members that have parallaxes with $\sigma_{\pi}<1$~mas,
I have used proper motions in a way that accounts for projection effects to
identify the individual groups based on their distinct kinematics.
Through this analysis, I have divided the members with parallaxes into 13 
groups, which are named after their associated dark clouds or their brightest 
stellar members. Some of those Taurus groups differ substantially 
from the groups produced by recent studies.
Those differences are explained in part by the fact that clustering analysis
based on proper motions or tangential velocities without correction for
projection effects is prone to breaking extended associations into small
fragments and assigning stars with different velocities to the same group.

\item
My survey for new members of the Taurus groups with Gaia DR3 has been sensitive
to stars with ages of $\lesssim20$~Myr. I find no evidence of a population of 
older stars (10--20~Myr) that is comoving with the Taurus groups, which is 
consistent with the results of a previous search with Gaia DR2 \citep{luh18}.

\item
Previous studies have identified several groups and associations
of young stars within a large volume surrounding the Taurus clouds 
\citep{mam07,mam16,bel17,oh17,luh18,liu20,liu21,gag20,ker21}. I have performed
a survey for young associations near Taurus using Gaia DR3, which has resulted
in 1378 candidate members of seven associations. A similar survey of
an additional association near Taurus, 32~Ori, was recently presented 
in \citet{luh22o}. The numbers of candidates in these associations range from
47 to 591, most of which have not been previously identified.
As found in previous work \citep{gag20}, my results illustrate that
surveys based on clustering of proper motions or tangential velocities often
identify only small fragments of associations.

\item
All of the associations near Taurus have histograms of spectral types
that peak near M5 ($\sim0.15$~$M_\odot$), indicating that they have IMFs
with similar characteristic masses to other nearby associations and
star-forming regions.

\item
For the nine largest Taurus groups and the neighboring associations, 
I have measured the offsets in absolute magnitudes of their sequences of 
low-mass stars from the median sequence of UCL/LCC in H-R diagrams.
Those offsets have been converted to ages by assuming that UCL/LCC
has an age of 20~Myr and that the luminosities fade at a rate given by
$\Delta$log~L/$\Delta$log~age$=-0.6$, which is typical of the predictions
of evolutionary models \citep{bar15,cho16,dot16,fei16}.
That method produces ages that are broadly consistent with the values
for the $\beta$~Pic moving group \citep[$\sim22$~Myr,][]{bin16} and the
Pleiades cluster \citep[$\sim120$~Myr,][]{sta98,dah15} based on
the lithium depletion boundary.
Eight Taurus groups have median ages between $\sim$1 and 3~Myr while the
remaining group, HD~28354, as an age of $\sim$6~Myr. The age uncertainties 
are large enough that most of the groups could be coeval. 
The age estimates for the associations near Taurus range from 13 to 56~Myr.

\item
I have used mid-IR photometry from WISE to search for IR excesses from
circumstellar disks among the candidate members of the associations near Taurus.
Disks are detected for 51 stars, 20 of which are reported for the first time
in this work. Most of the disk-bearing stars have not been previously
identified as members of associations.
I have classified the evolutionary stages of the disks using the sizes of
their mid-IR excesses. Roughly half of the disks are classified as
full, transitional, or evolved, making them relatively old examples of
primordial disks.

\item
I have calculated $UVW$ velocities for members of Taurus and the neighboring
associations that have measurements of proper motions, parallaxes, and
radial velocities. As found with the proper motion data, the associations
have velocities that are distinct from those of the Taurus groups.
There is also little spatial overlap between the Taurus groups and the
associations. 

\item
Previous studies have proposed that supernovae in UCL/LCC beginning 10--15~Myr
ago powered the expansion of the Local Bubble, sweeping up gas to form Taurus
and other nearby molecular clouds and triggering their star formation
\citep[e.g.,][]{zuc22}.
In that scenario, one might expect an age gradient among the Taurus groups
that corresponds to the crossing time of the expanding shell ($\sim$5~Myr).
A gradient of that kind is not evident when I use the median velocities of 
the groups to estimate their relative positions in the past, although it
might only be marginally detectable given the uncertainties of the group ages.
In addition, the interface between the interstellar medium and the Local 
Bubble may have been too clumpy and irregular for a gradient to appear.

\item
Some recent studies have contended that samples of older stars ($\gtrsim$10~Myr)
found in the vicinity of Taurus represent a distributed population produced
by an earlier epoch of star formation in the Taurus complex 
\citep{kra17,ker21,kro21}. I find that the older stars in question 
consist of a mixture of candidate members of the associations near Taurus 
and stars spanning a wide range of kinematics that do not comprise a coherent 
population and instead are likely to be young field stars.
These stars do not share the same motions as the Taurus groups, indicating
that they did not arise from the existing clouds.
In addition, very few of the older stars that have $UVW$ velocity
measurements trace back to birth sites that would have been near the Taurus
groups (or their progenitor gas), demonstrating that they have no relationship 
with the latter. The same is true for most of the full samples of candidates
that I have identified in the associations near Taurus.
The one exception is the small association containing V1362~Tau (53 stars), 
which has an age of 13~Myr and a projected birth site that was within
$\sim7$~pc from the L1527 group in Taurus.  V1362~Tau could represent the 
first group born in the Taurus complex, although a close approach of the 
Taurus clouds with an unrelated association is plausible as well. 
In either case, I find that a distributed population of older stars associated
with the Taurus clouds does not exist.

\item
My catalogs for the associations near Taurus include a few stars that are
well-studied or notable, such as the Herbig Ae/Be stars HD~35187, 
CQ~Tau, and MWC~758 (members of the HD~35187 association) and the 
planet-hosting stars V1298~Tau and V1069~Tau (members of the
93~Tau association).

\end{enumerate}

\acknowledgements
I thank Lee Hartmann for comments on the manuscript.
This work used data from the European Space Agency
(ESA) mission Gaia (\url{https://www.cosmos.esa.int/gaia}), processed by
the Gaia Data Processing and Analysis Consortium (DPAC,
\url{https://www.cosmos.esa.int/web/gaia/dpac/consortium}). Funding
for the DPAC has been provided by national institutions, in particular
the institutions participating in the Gaia Multilateral Agreement.
The IRTF is operated by the University of Hawaii under contract 80HQTR19D0030
with NASA. The Gemini data were obtained through program
GN-2019B-Q-222 (NOAO program 2019B-0124). The observations at the KPNO 2.1~m
telescope were performed through program 2010B-0530 at NOIRLab.
This work used data provided by the Astro Data Archive at NOIRLab.
NOIRLab is operated by the Association of Universities for
Research in Astronomy under a cooperative agreement with the NSF.
Gemini Observatory is a program of NSF's NOIRLab, which is managed by the
Association of Universities for Research in Astronomy (AURA) under a
cooperative agreement with the National Science Foundation on behalf of the 
Gemini Observatory partnership: the National Science Foundation (United States),
National Research Council (Canada), Agencia Nacional de Investigaci\'{o}n y 
Desarrollo (Chile), Ministerio de Ciencia, Tecnolog\'{i}a e Innovaci\'{o}n 
(Argentina), Minist\'{e}rio da Ci\^{e}ncia, Tecnologia, Inova\c{c}\~{o}es e 
Comunica\c{c}\~{o}es (Brazil), and Korea Astronomy and Space Science Institute 
(Republic of Korea).
2MASS is a joint project of the University of Massachusetts and IPAC
at Caltech, funded by NASA and the NSF.
WISE is a joint project of the University of California, Los Angeles,
and the JPL/Caltech, funded by NASA. This work used data from the
NASA/IPAC Infrared Science Archive, operated by JPL under contract
with NASA, and the VizieR catalog access tool and the SIMBAD database,
both operated at CDS, Strasbourg, France.
LAMOST is a National Major Scientific Project built by the Chinese Academy of Sciences. Funding for the project has been provided by the National Development and Reform Commission. LAMOST is operated and managed by the National Astronomical Observatories, Chinese Academy of Sciences.
Funding for the Sloan Digital Sky Survey IV has been provided by the 
Alfred P. Sloan Foundation, the U.S. Department of Energy Office of 
Science, and the Participating Institutions. 
SDSS-IV acknowledges support and resources from the Center for High 
Performance Computing  at the University of Utah. The SDSS 
website is www.sdss.org.  SDSS-IV is managed by the 
Astrophysical Research Consortium for the Participating Institutions 
of the SDSS Collaboration including the Brazilian Participation Group, 
the Carnegie Institution for Science, Carnegie Mellon University, Center for 
Astrophysics | Harvard \& Smithsonian, the Chilean Participation 
Group, the French Participation Group, Instituto de Astrof\'isica de 
Canarias, The Johns Hopkins University, Kavli Institute for the 
Physics and Mathematics of the Universe (IPMU) / University of 
Tokyo, the Korean Participation Group, Lawrence Berkeley National Laboratory, 
Leibniz Institut f\"ur Astrophysik Potsdam (AIP),  Max-Planck-Institut 
f\"ur Astronomie (MPIA Heidelberg), Max-Planck-Institut f\"ur 
Astrophysik (MPA Garching), Max-Planck-Institut f\"ur 
Extraterrestrische Physik (MPE), National Astronomical Observatories of 
China, New Mexico State University, New York University, University of 
Notre Dame, Observat\'ario Nacional / MCTI, The Ohio State 
University, Pennsylvania State University, Shanghai 
Astronomical Observatory, United Kingdom Participation Group, 
Universidad Nacional Aut\'onoma de M\'exico, University of Arizona, 
University of Colorado Boulder, University of Oxford, University of 
Portsmouth, University of Utah, University of Virginia, University 
of Washington, University of Wisconsin, Vanderbilt University, 
and Yale University.
The Center for Exoplanets and Habitable Worlds is supported by the
Pennsylvania State University, the Eberly College of Science, and the
Pennsylvania Space Grant Consortium.

\clearpage

\begin{deluxetable}{rllll}
\tabletypesize{\scriptsize}
\tablewidth{0pt}
\tablecaption{New Spectral and Disk Classifications of Taurus Members\label{tab:spec}}
\tablehead{
\colhead{Gaia DR3} &
\colhead{Other Name} &
\colhead{Spectral Type} &
\colhead{Telescope/Instrument} & 
\colhead{Disk Type}}
\startdata
3420750548559422592 & PSO J079.3986+26.2455 & M6.5 & IRTF/SpeX & full \\
3414676232147787136 & \nodata & \nodata & \nodata & III \\
3419115132386033280 & \nodata & M3.75 & IRTF/SpeX & \nodata \\
145200960104259200 & \nodata & M0.5 & LAMOST & III \\
3418846267435680512 & \nodata & M5.25 & LAMOST & full \\
3446890376655192832 & \nodata & M5.5 & LAMOST & III \\
3446890411014932224 & \nodata & M5.25 & LAMOST & III \\
3446722593755425024 & PW Aur & M3 & LAMOST & full \\
180149418232233472 & \nodata & M5.5 & LAMOST & full \\
156207518176564864 & \nodata & M4.75 & LAMOST & full \\
157247965413620224 & \nodata & M5.75 & LAMOST & full \\
156162674425653248 & \nodata & M1 & KPNO~2.1~m/GoldCam & full? \\
3420824426291884672 & \nodata & M0.5 & KPNO~2.1~m/GoldCam & full\\
3421359544857244160 & \nodata & M5.5 & LAMOST & III 
\enddata
\end{deluxetable}

\clearpage

\begin{deluxetable}{ll}
\tabletypesize{\scriptsize}
\tablewidth{0pt}
\tablecaption{Adopted Members of Taurus\label{tab:mem}}
\tablehead{
\colhead{Column Label} &
\colhead{Description}}
\startdata
GaiaDR3 & Gaia DR3 source name\\
2MASS & 2MASS source name\\
UGCS & UKIDSS Galactic Clusters Survey source name\tablenotemark{a}\\
Name & Other source name\\
RAdeg & Right ascension (ICRS)\\
DEdeg & Declination (ICRS)\\
Ref-Pos & Reference for right ascension and declination\tablenotemark{b}\\
SpType & Spectral type \\
r\_SpType & Spectral type reference\tablenotemark{c}\\
Adopt & Adopted spectral type \\
pmRA & Gaia DR3 proper motion in right ascension\tablenotemark{d}\\
e\_pmRA & Error in pmRA\tablenotemark{d}\\
pmDec & Gaia DR3 proper motion in declination\tablenotemark{d}\\
e\_pmDec & Error in pmDec\tablenotemark{d}\\
plx & Gaia DR3 parallax\tablenotemark{d}\\
e\_plx & Error in plx\tablenotemark{d}\\
r\_med\_geo & Median of the geometric distance posterior \citep{bai21}\tablenotemark{d}\\
r\_lo\_geo & 16th percentile of the geometric distance posterior \citep{bai21}\tablenotemark{d}\\
r\_hi\_geo & 84th percentile of the geometric distance posterior \citep{bai21}\tablenotemark{d}\\
RVel & Radial velocity \\
e\_RVel & Error in RVel \\
vscatter & {\tt VSCATTER} for radial velocity from SDSS-IV DR17 \citep{apo17}\\
r\_RVel & Radial velocity reference\tablenotemark{e}\\
U & $U$ component of space velocity \\
e\_U & Error in U \\
V & $V$ component of space velocity \\
e\_V & Error in V \\
W & $W$ component of space velocity \\
e\_W & Error in W \\
Gmag & Gaia DR3 $G$ magnitude\\
e\_Gmag & Error in Gmag \\
GBPmag & Gaia DR3 $G_{\rm BP}$ magnitude\\
e\_GBPmag & Error in GBPmag \\
GRPmag & Gaia DR3 $G_{\rm RP}$ magnitude\\
e\_GRPmag & Error in GRPmag \\
RUWE & Gaia DR3 renormalized unit weight error\\
outlier & Astrometric outlier\tablenotemark{f}\\
group & Taurus group based on Gaia DR3 astrometry
\enddata
\tablenotetext{a}{Based on coordinates from Data Release 10 of the UKIDSS
Galactic Clusters Survey for stars with $K_s>10$ from 2MASS.}
\tablenotetext{b}{Sources of the right ascension and declination are
are the 2MASS Point Source Catalog, Gaia DR3 (Epoch 2016.0), UKIDSS Data 
Release 10, and images from the Spitzer Space Telescope \citep{luh10}.}
\tablenotetext{c}{
(1) \citet{esp19};
(2) \citet{luh09tau};
(3) \citet{whi04};
(4) \citet{dop05};
(5) \citet{pra09};
(6) \citet{esp14};
(7) \citet{ngu12};
(8) \citet{wic96};
(9) \citet{luh17};
(10) \citet{esp17};
(11) \citet{her86};
(12) \citet{her14};
(13) \citet{her77};
(14) \citet{con10};
(15) \citet{tor95};
(16) \citet{sch09};
(17) \citet{reb10};
(18) \citet{luh04};
(19) \citet{har94};
(20) \citet{wel95};
(21) \citet{coh79};
(22) \citet{whi03};
(23) \citet{bri98};
(24) \citet{str94};
(25) \citet{har03};
(26) \citet{sce08};
(27) \citet{bri93};
(28) \citet{bri02};
(29) \citet{gui06};
(30) \citet{luh06tau1};
(31) \citet{lr98};
(32) \citet{luh03};
(33) \citet{luh98vx};
(34) \citet{mar01};
(35) \citet{pat93};
(36) \citet{ken98};
(37) \citet{bec07};
(38) \citet{luh99};
(39) \citet{mun83};
(40) \citet{duc99};
(41) \citet{ken90};
(42) \citet{cal04};
(43) \citet{mar99};
(44) \citet{luh06tau2};
(45) \citet{abe14};
(46) \citet{zha18};
(47) \citet{luh09fu};
(48) \citet{cie12};
(49) \citet{ken94};
(50) \citet{har91};
(51) \citet{wic00};
(52) \citet{whi01};
(53) \citet{sle06};
(54) \citet{bon14};
(55) \citet{giz99};
(56) \citet{abt04};
(57) \citet{mon98};
(58) \citet{her88};
(59) \citet{muz03};
(60) \citet{kra09};
(61) \citet{wal88};
(62) \citet{whi99};
(63) \citet{pra02};
(64) \citet{rei99};
(65) \citet{wal03};
(66) \citet{cow72};
(67) \citet{mar94};
(68) \citet{gom92};
(69) this work;
(70) \citet{kra17};
(71) \citet{her08};
(72) \citet{nes95};
(73) \citet{mar00};
(74) \citet{bow15};
(75) \citet{bri99};
(76) \citet{luh18};
(77) \citet{rac68};
(78) \citet{har95};
(79) \citet{bos37};
(80) \citet{ste01};
(81) \citet{mal98};
(82) \citet{li98};
(83) \citet{liu21};
(84) \citet{fin10}.}
\tablenotetext{d}{Proper motions, parallaxes, and distances for
LkCa~3~A, V410~Anon~25, XZ~Tau~A, LkHa332/G1, and V807~Tau are from
radio interferometry \citep{gal18}.}
\tablenotetext{e}{
(1) Gaia DR3;
(2) \citet{apo17};
(3) \citet{gon06};
(4) \citet{tor13};
(5) \citet{ngu12};
(6) \citet{kro21};
(7) \citet{kou19};
(8) \citet{har86};
(9) \citet{zha21} and LAMOST DR7;
(10) \citet{bas95};
(11) \citet{whi03};
(12) \citet{rei90};
(13) \citet{kra17};
(14) \citet{muz03};
(15) \citet{mat97}.}
\tablenotetext{f}{
* = outlier in distance or proper motion offset in
Figures~\ref{fig:pp1}--\ref{fig:pp4}.}
\tablecomments{The table is available in a machine-readable form.}
\end{deluxetable}

\clearpage

\begin{deluxetable}{ll}
\tabletypesize{\scriptsize}
\tablewidth{0pt}
\tablecaption{Candidate Members of Taurus that Lack Spectroscopy\label{tab:cand}}
\tablehead{
\colhead{Column Label} &
\colhead{Description}}
\startdata
GaiaDR3 & Gaia DR3 source name\\
2MASS & 2MASS source name\\
WISEA & AllWISE source name\tablenotemark{a}\\
Name & Other source name \\
RAdeg & Gaia DR3 right ascension (ICRS at Epoch 2016.0)\\
DEdeg & Gaia DR3 declination (ICRS at Epoch 2016.0)\\
pmRA & Gaia DR3 proper motion in right ascension\\
e\_pmRA & Error in pmRA\\
pmDec & Gaia DR3 proper motion in declination\\
e\_pmDec & Error in pmDec\\
plx & Gaia DR3 parallax\\
e\_plx & Error in plx \\
r\_med\_geo & Median of the geometric distance posterior \citep{bai21}\\
r\_lo\_geo & 16th percentile of the geometric distance posterior \citep{bai21}\\
r\_hi\_geo & 84th percentile of the geometric distance posterior \citep{bai21}\\
RVel & Gaia DR3 radial velocity \\
e\_RVel & Error in RVel \\
U & $U$ component of space velocity \\
e\_U & Error in U \\
V & $V$ component of space velocity \\
e\_V & Error in V \\
W & $W$ component of space velocity \\
e\_W & Error in W \\
Gmag & Gaia DR3 $G$ magnitude\\
e\_Gmag & Error in Gmag \\
GBPmag & Gaia DR3 $G_{\rm BP}$ magnitude\\
e\_GBPmag & Error in GBPmag \\
GRPmag & Gaia DR3 $G_{\rm RP}$ magnitude\\
e\_GRPmag & Error in GRPmag \\
RUWE & Gaia DR3 renormalized unit weight error\\
Jmag & 2MASS $J$ magnitude \\
e\_Jmag & Error in Jmag \\
Hmag & 2MASS $H$ magnitude \\
e\_Hmag & Error in Hmag \\
Ksmag & 2MASS $K_s$ magnitude \\
e\_Ksmag & Error in Ksmag \\
W1mag & WISE W1 magnitude \\
e\_W1mag & Error in W1mag \\
W2mag & WISE W2 magnitude \\
e\_W2mag & Error in W2mag \\
W3mag & WISE W3 magnitude \\
e\_W3mag & Error in W3mag \\
f\_W3mag & Flag on W3mag\tablenotemark{b} \\
W4mag & WISE W4 magnitude \\
e\_W4mag & Error in W4mag \\
f\_W4mag & Flag on W4mag\tablenotemark{b} \\
DiskType & Disk type\\
group & Taurus group based on Gaia DR3 astrometry
\enddata
\tablenotetext{a}{The WISE source name for Gaia DR3156162678718169344 
is from the WISE All-Sky Catalog.}
\tablenotetext{b}{nodet = nondetection; false = detection from
WISE appears to be false or unreliable based on visual inspection.}
\tablecomments{The table is available in a machine-readable form.}
\end{deluxetable}

\clearpage

\LongTables

\begin{deluxetable}{ll}
\tabletypesize{\scriptsize}
\tablewidth{0pt}
\tablecaption{Candidate Members of Associations Near Taurus\label{tab:groups}}
\tablehead{
\colhead{Column Label} &
\colhead{Description}}
\startdata
GaiaDR3 & Gaia DR3 source name\\
Name & Other source name\\
RAdeg & Gaia DR3 right ascension (ICRS at Epoch 2016.0)\\
DEdeg & Gaia DR3 declination (ICRS at Epoch 2016.0)\\
SpType & Spectral type \\
r\_SpType & Spectral type reference\tablenotemark{a}\\
Adopt & Adopted spectral type \\
pmRA & Gaia DR3 proper motion in right ascension\\
e\_pmRA & Error in pmRA\\
pmDec & Gaia DR3 proper motion in declination\\
e\_pmDec & Error in pmDec\\
plx & Gaia DR3 parallax\\
e\_plx & Error in plx\\
r\_med\_geo & Median of the geometric distance posterior \citep{bai21}\\
r\_lo\_geo & 16th percentile of the geometric distance posterior \citep{bai21}\\
r\_hi\_geo & 84th percentile of the geometric distance posterior \citep{bai21}\\
RVel & Radial velocity \\
e\_RVel & Error in RVel \\
vscatter & {\tt VSCATTER} for radial velocity from SDSS-IV DR17 \citep{apo17}\\
r\_RVel & Radial velocity reference\tablenotemark{b}\\
U & $U$ component of space velocity \\
e\_U & Error in U \\
V & $V$ component of space velocity \\
e\_V & Error in V \\
W & $W$ component of space velocity \\
e\_W & Error in W \\
Gmag & Gaia DR3 $G$ magnitude\\
e\_Gmag & Error in Gmag \\
GBPmag & Gaia DR3 $G_{\rm BP}$ magnitude\\
e\_GBPmag & Error in GBPmag \\
GRPmag & Gaia DR3 $G_{\rm RP}$ magnitude\\
e\_GRPmag & Error in GRPmag \\
RUWE & Gaia DR3 renormalized unit weight error\\
2m & Closest 2MASS source within $3\arcsec$ \\
2msep & Angular separation between Gaia DR3 (epoch 2000) and 2MASS \\
2mclosest & Is this Gaia source the closest match for the 2MASS source? \\
wise & Closest WISE source within $3\arcsec$\tablenotemark{c} \\
wisesep & Angular separation between Gaia DR3 (epoch 2010.5) and WISE \\
wiseclosest & Is this Gaia source the closest match for the WISE source?\\
Jmag & 2MASS $J$ magnitude \\
e\_Jmag & Error in Jmag \\
Hmag & 2MASS $H$ magnitude \\
e\_Hmag & Error in Hmag \\
Ksmag & 2MASS $K_s$ magnitude \\
e\_Ksmag & Error in Ksmag \\
W1mag & WISE W1 magnitude \\
e\_W1mag & Error in W1mag \\
W2mag & WISE W2 magnitude \\
e\_W2mag & Error in W2mag \\
W3mag & WISE W3 magnitude \\
e\_W3mag & Error in W3mag \\
f\_W3mag & Flag on W3mag\tablenotemark{d} \\
W4mag & WISE W4 magnitude \\
e\_W4mag & Error in W4mag \\
f\_W4mag & Flag on W4mag\tablenotemark{d} \\
ExcW2 & Excess present in W2? \\
ExcW3 & Excess present in W3? \\
ExcW4 & Excess present in W4? \\
DiskType & Disk Type \\
association & Association
\enddata
\tablenotetext{a}{
(1) \citet{liu21};
(2) measured in this work with LAMOST DR7 data;
(3) measured in this work with GMOS data;
(4) \citet{sle06};
(5) \citet{ria06};
(6) \citet{can93};
(7) \citet{esp19};
(8) \citet{nes95};
(9) \citet{wic96};
(10) \citet{whi07};
(11) \citet{wal88};
(12) \citet{pat93};
(13) \citet{har03};
(14) \citet{her14};
(15) measured in this work with Red Channel data;
(16) \citet{kra17};
(17) \citet{can49};
(18) \citet{fin10};
(19) \citet{luh17};
(20) \citet{zha18};
(21) measured in this work with SpeX data from \citet{zha18};
(22) \citet{gom92};
(23) \citet{bri98};
(24) \citet{rei99};
(25) \citet{esp17};
(26) \citet{esp14};
(27) \citet{luh06tau2};
(28) \citet{cow69};
(29) \citet{moo13};
(30) \citet{neu95};
(31) \citet{mag97};
(32) \citet{giz99};
(33) \citet{man17};
(34) \citet{li98};
(35) \citet{alc96};
(36) \citet{alc00};
(37) \citet{bia12};
(38) \citet{gre99};
(39) \citet{abt08};
(40) \citet{mcc59};
(41) \citet{cow72};
(42) \citet{bid88};
(43) \citet{bin15};
(44) \citet{vie03};
(45) \citet{mor01};
(46) \citet{gag15};
(47) \citet{cow68};
(48) \citet{wil52};
(49) \citet{zic05};
(50) \citet{hou99};
(51) \citet{pau01};
(52) \citet{bri19}.}
\tablenotetext{b}{
(1) Gaia DR3;
(2) \citet{apo17};
(3) \citet{ngu12};
(4) \citet{sou18};
(5) \citet{wal88};
(6) \citet{kra17};
(7) \citet{mer09};
(8) \citet{zha21} and LAMOST DR7;
(9) \citet{rei99};
(10) \citet{gon06};
(11) Gaia DR2;
(12) \citet{bia12};
(13) \citet{gal3};
(14) \citet{gre99}.}
\tablenotetext{c}{Source name from AllWISE Source Catalog, AllWISE Reject
Catalog, or WISE All-Sky Source Catalog.}
\tablenotetext{d}{nodet = nondetection; false = detection from
WISE catalog appears to be false or unreliable based on visual inspection.}
\tablecomments{The table is available in a machine-readable form.}
\end{deluxetable}

\clearpage

\begin{deluxetable}{cc}
\tabletypesize{\scriptsize}
\tablewidth{0pt}
\tablecaption{Fit to the Median of the Sequence of Low-mass Stars in
UCL/LCC\label{tab:fit}}
\tablehead{
\colhead{$G_{\rm BP}-G_{\rm RP}$} &
\colhead{$M_{G_{\rm RP}}$}} 
\startdata
      1.4 &   5.18 \\
      1.5 &   5.37 \\
      1.6 &   5.56 \\
      1.7 &   5.75 \\
      1.8 &   5.94 \\
      1.9 &   6.13 \\
      2.0 &   6.32 \\
      2.1 &   6.51 \\
      2.2 &   6.70 \\
      2.3 &   6.85 \\
      2.4 &   7.00 \\
      2.5 &   7.18 \\
      2.6 &   7.40 \\
      2.7 &   7.65 \\
      2.8 &   7.92
\enddata
\tablecomments{The fit can be transformed to spectral types and
other bands using the typical intrinsic colors of young stars
from \citet{luh22sc}.}
\end{deluxetable}

\clearpage

\begin{deluxetable}{lrrr}
\tabletypesize{\scriptsize}
\tablewidth{0pt}
\tablecaption{Median Absolute Magnitude Offsets and Relative Ages of Taurus Groups and Neighboring Associations\label{tab:ages}}
\tablehead{
\colhead{Group/Association} &
\colhead{$\Delta M_{G_{\rm RP}/K}$\tablenotemark{a}} &
\colhead{N$_*$\tablenotemark{b}} &
\colhead{Age\tablenotemark{c}}\\
\colhead{} &
\colhead{(mag)} &
\colhead{} &
\colhead{(Myr)}}
\startdata
\cutinhead{Taurus Groups}
L1527 & $-$2.32$\pm$0.70 & 5 & 0.6$^{+1.1}_{-0.4}$ \\
L1495/B209 & $-$1.68$\pm$0.87 & 19 & 1.5$^{+4.3}_{-1.1}$ \\
L1521/B213 & $-$1.52$\pm$1.01 & 8 & 1.9$^{+7.3}_{-1.5}$ \\
L1536 & $-$1.48$\pm$0.70 & 15 & 2.1$^{+4.0}_{-1.4}$ \\
L1517 & $-$1.35$\pm$0.60 & 21 & 2.5$^{+3.8}_{-1.5}$ \\
L1524/L1529/B215 & $-$1.32$\pm$0.93 & 10 & 2.6$^{+8.4}_{-2.0}$ \\
L1551 & $-$1.30$\pm$0.76 & 17 & 2.7$^{+6.0}_{-1.9}$ \\
L1544 & $-$1.11$\pm$0.87 & 6 & 3.6$^{+10.1}_{-2.7}$ \\
HD28354 & $-$0.78$\pm$0.36 & 7 & 6.0$^{+4.5}_{-2.6}$ \\
\cutinhead{Associations and Clusters Near Taurus}
V1362 Tau & $-$0.30$\pm$0.26 & 19 & 13$\pm$5 \\
HD35187 & $-$0.08$\pm$0.15 & 16 & 18$\pm$4 \\
HD33413 & $-$0.05$\pm$0.12 & 16 & 19$\pm$3 \\
HD284346 & 0.20$\pm$0.10 & 14 & 27$\pm$4 \\
93 Tau & 0.36$\pm$0.24 & 40 & 35$\pm$13 \\
69 Ori & 0.63$\pm$0.13 & 137 & 53$\pm$11 \\
$\mu$ Tau & 0.67$\pm$0.11 & 69 & 56$\pm$9 \\
$\alpha$ Per & 0.73$\pm$0.11 & 116 & 61$\pm$11 \\
Pleiades & 1.19$\pm$0.12 & 189 & 124$\pm$24
\enddata
\tablenotetext{a}{Median of the offsets from the median sequence for
UCL/LCC in $M_K$ (Taurus groups) or $M_{G_{\rm RP}}$ (other associations)
for low-mass stars.} 
\tablenotetext{b}{Number of low-mass stars used in calculation of 
$\Delta M_{G_{\rm RP}/K}$.}
\tablenotetext{c}{Calculated from $\Delta M_{G_{\rm RP}/K}$ assuming 
an age of 20~Myr for UCL/LCC and $\Delta$log~L/$\Delta$log~age$=-0.6$.}
\end{deluxetable}

\clearpage

\begin{deluxetable}{lcrrrrrrrrrr}
\tabletypesize{\scriptsize}
\tablewidth{0pt}
\tablecaption{Median Distances and Kinematics for Taurus Groups and
Neighboring Associations\label{tab:uvw}}
\tablehead{
\colhead{Group} &
\colhead{Distance\tablenotemark{a}} &
\colhead{$\mu_\alpha$\tablenotemark{a}} &
\colhead{$\mu_\delta$\tablenotemark{a}} &
\colhead{N$_*$} &
\colhead{$U$\tablenotemark{b}} &
\colhead{$V$\tablenotemark{b}} &
\colhead{$W$\tablenotemark{b}} &
\colhead{$\sigma_U$\tablenotemark{b}} &
\colhead{$\sigma_V$\tablenotemark{b}} &
\colhead{$\sigma_W$\tablenotemark{b}} &
\colhead{N$_*$}\\
\colhead{} &
\colhead{(pc)} &
\multicolumn{2}{c}{(mas~yr$^{-1}$)} &
\colhead{} &
\multicolumn{3}{c}{(km~s$^{-1}$)} &
\multicolumn{3}{c}{(km~s$^{-1}$)} &
\colhead{}}
\startdata
\cutinhead{Taurus Groups}
L1495/B209 & 130 & 8.8 & $-$25.2 & 66 & $-$16.4 & $-$11.9 & $-$10.9 & 1.9 & 1.1 & 1.1 & 47\\
B209N & 159 & 12.1 & $-$18.1 & 5 & $-$18.7 & $-$12.4 & $-$7.4 & \nodata & \nodata & \nodata & 3\\
L1489/L1498 & 149 & 14.2 & $-$18.8 & 4 & $-$17.9 & $-$13.7 & $-$7.3 & \nodata & \nodata & \nodata & 1\\
L1521/B213 & 158 & 11.3 & $-$17.7 & 26 & $-$17.7 & $-$12.8 & $-$7.0 & 0.8 & 0.7 & 0.7 & 16\\
HD28354 & 129 & 8.9 & $-$26.7 & 16 & $-$16.2 & $-$13.9 & $-$10.7 & 1.7 & 0.8 & 0.8 & 9\\
L1524/L1529/B215 & 128 & 7.0 & $-$21.3 & 59 & $-$15.8 & $-$10.9 & $-$9.6 & 1.1 & 0.7 & 0.7 & 37\\
L1517 & 156 & 4.4 & $-$24.7 & 57 & $-$16.0 & $-$14.5 & $-$10.7 & 2.4 & 0.8 & 0.8 & 26\\
L1527 & 140 & 5.4 & $-$20.2 & 25 & $-$15.9 & $-$11.5 & $-$9.7 & 1.1 & 0.8 & 0.8 & 18\\
L1544 & 168 & 3.0 & $-$17.3 & 14 & $-$18.2 & $-$12.2 & $-$9.3 & 0.7 & 0.8 & 0.8 & 4\\
L1536 & 161 & 10.3 & $-$17.0 & 35 & $-$16.5 & $-$13.8 & $-$7.0 & 1.0 & 1.1 & 1.1 & 23\\
L1551 & 144 & 12.1 & $-$19.0 & 54 & $-$16.6 & $-$15.2 & $-$7.6 & 2.0 & 0.9 & 0.9 & 33\\
T Tau & 144 & 7.0 & $-$12.3 & 5 & $-$18.0 & $-$8.4 & $-$8.2 & \nodata & \nodata & \nodata & 3\\
L1558 & 196 & 5.0 & $-$13.3 & 7 & $-$19.3 & $-$13.8 & $-$10.4 & 0.5 & 0.3 & 0.3 & 4\\
\cutinhead{Associations Near Taurus}
93 Tau & 119 & 0.0 & $-$14.5 & 190 & $-$13.4 & $-$6.4 & $-$9.8 & 2.4 & 0.6 & 0.6 & 69\\
HD284346 & 192 & 4.8 & $-$6.2 & 78 & $-$14.6 & $-$5.9 & $-$5.8 & 2.9 & 0.2 & 0.2 & 15\\
HD33413 & 173 & 3.0 & $-$19.7 & 47 & $-$16.8 & $-$15.1 & $-$9.2 & 2.3 & 0.7 & 0.7 & 9\\
69 Ori & 199 & 7.7 & $-$19.3 & 591 & $-$17.0 & $-$22.6 & $-$3.7 & 2.5 & 0.7 & 0.7 & 78\\
V1362 Tau & 176 & 1.3 & $-$18.2 & 53 & $-$18.8 & $-$13.7 & $-$9.2 & 1.4 & 0.5 & 0.5 & 14\\
HD35187 & 160 & 3.9 & $-$25.8 & 65 & $-$22.4 & $-$17.9 & $-$10.7 & 5.6 & 0.7 & 0.7 & 7\\
$\mu$ Tau & 156 & 22.2 & $-$22.1 & 354 & $-$15.4 & $-$23.4 & $-$7.7 & 1.9 & 1.0 & 1.0 & 72
\enddata
\tablenotetext{a}{Median value for members that have Gaia DR3 proper motions
and parallaxes and are not outliers in Figures~\ref{fig:pp1}--\ref{fig:pp4}.}
\tablenotetext{b}{Median value for members that have Gaia DR3 proper motions
and parallaxes, are not outliers in Figures~\ref{fig:pp1}--\ref{fig:pp4}, 
have radial velocity measurements, and have {\tt VSCATTER}$<3$~km~s$^{-1}$ from 
\citet{apo17} when available.}
\end{deluxetable}

\clearpage

\begin{deluxetable}{cl}
\tabletypesize{\scriptsize}
\tablewidth{0pt}
\tablecaption{Group Classifications for Stars in Components from
\citet{kro21}\label{tab:kro}}
\tablehead{
\colhead{Component from} &
\colhead{Group Classifications}\\
\colhead{\citet{kro21}} &
\colhead{from this work\tablenotemark{a}}}
\startdata
C1 & L1551 (38), T Tau (5), field star (1) \\
C2 & L1495/B209 (55) \\
C3 & L1517 (28), V1362~Tau (3), HD~33413 (1), field star (4) \\
C4 & L1517 (25), field star (1) \\
C5 & L1536 (32) \\
C6 & L1524/L1529/B215 (52), HD~28354 (14), L1495/B209 (2), L1527 (1), L1536 (1), 93~Tau (1),\\
 & Taurus member not assigned to a group (1) \\
C7 & L1527 (22) \\
C8 & L1521/B213 (20), L1524/L1529/B215 (1) \\
C9 & 32 Ori (4), field star (2) \\
C10 & 32 Ori (3) \\
D1 & L1544 (12), V1362 Tau (8), HD~35187 (2), field star (2) \\
D2 & L1551 (20), L1558 (9), 32 Ori (1), 93~Tau (6), field star (19) \\
D3 & 93~Tau (16), L1524/L1529/B215 (3), L1495/B209 (2), field star (4) \\
D4 & L1495/B209 (10), L1524/L1529/B215 (10), L1521/B213 (8), L1536 (8), L1489/L1498 (6), B209N (6),\\
 & 93~Tau (4), 32 Ori (3), HD~28354 (2), Taurus member not assigned to a group (2), field star (19)
\enddata
\tablenotetext{a}{Numbers of stars are indicated in parentheses.}
\end{deluxetable}

\clearpage

\end{document}